\documentclass[aps,showpacs,nofootinbib,twocolumn,superscriptaddress,floatfix]{revtex4-1}
\usepackage{color}
\usepackage{amsmath}
\usepackage{amsfonts}
\usepackage{amssymb}
\usepackage{enumerate}
\usepackage{slashed}
\usepackage{bm}
\usepackage[utf8]{inputenc}
\usepackage{datetime}
\usepackage{mciteplus}
\usepackage{amsmath}
\usepackage{graphicx}
\usepackage{esint}
\usepackage{ulem}

\usepackage{dsfont}
\definecolor{darkgreen}{rgb}{0,0.65,0}

\definecolor{green}{rgb}{0,.5,0}

\definecolor{orange}{rgb}{1,0.5,0}

\begin{document}

\title{Reconstructing parton densities at large fractional momenta}

\author{Alessandro Bacchetta}
\email{alessandro.bacchetta@unipv.it}
\affiliation{Dipartimento di Fisica, Universit\`a degli Studi di Pavia, via Bassi 6, I-27100 Pavia, Italy, and}
\affiliation{INFN Sezione di Pavia, via Bassi 6, I-27100 Pavia, Italy}

\author{Marco Radici}
\email{marco.radici@pv.infn.it}
\affiliation{INFN Sezione di Pavia, via Bassi 6, I-27100 Pavia, Italy}

\author{Barbara Pasquini}
\email{barbara.pasquini@unipv.it}
\affiliation{Dipartimento di Fisica, Universit\`a degli Studi di Pavia, via Bassi 6, I-27100 Pavia, Italy, and}
\affiliation{INFN Sezione di Pavia, via Bassi 6, I-27100 Pavia, Italy}

\author{Xiaonu Xiong}
\email{xiaonu.xiong@pv.infn.it}
\affiliation{INFN Sezione di Pavia, via Bassi 6, I-27100 Pavia, Italy}

\begin{abstract}
Parton distribution functions (PDFs) are nonperturbative objects defined by nonlocal light-cone correlations. 
They cannot be computed directly from Quantum Chromodynamics (QCD). Using a standard lattice QCD approach, it is possible to compute moments of PDFs, which are matrix elements of local operators. Recently, an alternative approach has been proposed, based on the introduction of quasi-parton distribution functions (quasi-PDFs), which are matrix elements of equal-time spatial correlations and hence calculable on lattice. Quasi-PDFs approach standard PDFs in the limit of very large longitudinal proton momenta $P^z$. This limit is not attainable in lattice simulations, and quasi-PDFs fail to reproduce PDFs at high fractional longitudinal momenta. In this paper, we propose a method to improve the reconstruction of PDFs by combining information from quasi-PDFs and from the Mellin moments of regular PDFs. We test our method using the diquark spectator model for up and down valence distributions of both unpolarized and helicity PDFs. In the future, the method can be used to produce PDFs entirely based on lattice QCD results. 
\end{abstract}

\date{\today, \currenttime}


\maketitle

%%%%%%%%%%%%%%%%%%%%%%%%%%%%%%%%%%%%%%

\section{introduction}
\label{sec:intro}

Parton distribution functions (PDFs) describe combinations of number densities of quarks and gluons in a fast-moving hadron. They depend on the fractional momentum $x$ carried by partons moving collinearly with the parent hadron. PDFs can be defined in field theory as hadronic matrix elements of correlation operators that are nonlocal on the light-cone~\cite{Furmanski:1981cw}. They are essentially nonperturbative objects, hence they cannot be computed from first principles in QCD using perturbative techniques. They can be isolated through appropriate factorization theorems~\cite{Collins:1989gx}; consequently, they also depend on the factorization scale, which usually coincides with the hard scale $Q^2$ of the process at hand. In these conditions, PDFs are extracted from global fits of experimental data within the collinear factorization framework (see Ref.~\cite{Rojo:2015acz} and references therein). The determination of their theoretical uncertainty is of fundamental importance for the interpretation of measurements at any hadronic collider, in particular at LHC when searching for effects induced from new physics beyond the Standard Model~\cite{Butterworth:2015oua}.

Lattice QCD is at present the most successful approach to solve QCD in the nonperturbative regime. However, PDFs cannot be directly computed in lattice QCD because the light-cone separation becomes complex in Euclidean space-time. In lattice QCD, only Mellin moments of PDFs can be computed because they reduce to hadronic matrix elements of local operators~\cite{Gockeler:2004wp,Deka:2008xr}. In practice, only few Mellin moments are available because of the limited computational power and of the operator mixing between higher and lower moments (see Ref.~\cite{Detmold:2005gg} for a discussion on how to overcome these problems). 

Recently, a new approach denoted as Large Momentum Effective field Theory (LaMET) has been proposed to approximate PDFs on lattice in terms of the so-called quasi-PDFs~\cite{Ji:2013dva}. Quasi-PDFs are obtained from hadronic matrix elements of equal-time spatial correlation operators. As such, they do not depend on time and can be calculated on an Euclidean lattice. Quasi-PDFs depend on the parton momentum fraction of the longitudinal hadron momentum $P^z$: $x=k^z/P^z$, where $k^z$ is the longitudinal momentum of parton. They reduce to the usual PDFs in the limit $P^z \to \infty$. In reality, only finite values of the hadron momentum can be sampled on lattice and a suitable factorization theorem needs to be derived to connect quasi-PDFs to PDFs. Since both functions have the same infrared behaviour, the connection is established by perturbatively computing matching coefficients. The latter ones are currently available up to one loop~\cite{Xiong:2013bka,Ma:2014jla} also for quasi generalized parton distributions~\cite{Ji:2015qla,Xiong:2015nua}. The renormalization of quasi-PDFs has been computed up to two loops~\cite{Ji:2015jwa}. 

Lattice calculations of quasi-PDFs have already been produced, but only for proton's momenta $P^z$ of the order of the proton mass~\cite{Alexandrou:2015rja,Lin:2014zya}. Larger values of $P^z$ are currently not reachable because the computational effort is too demanding. Hence, present results are plagued by contributions from higher twists and target mass corrections, that have been addressed for the first time in Ref.~\cite{Chen:2016utp}. 

Model calculations of quasi-PDFs are available in the framework of the diquark spectator approximation~\cite{Gamberg:2014zwa}. By comparing these results with the model expressions of standard PDFs in the same context~\cite{Bacchetta:2008af}, the authors of Ref.~\cite{Gamberg:2014zwa} find that for moderate $P^z$ the quasi-PDFs are a good approximation to PDFs only for intermediate $0.1 \lesssim x \lesssim 0.4$. Strong deviations are reported for large $x \to 1$, also because the support of quasi-PDFs is not restricted to the interval $[0,1]$~\cite{Xiong:2013bka,Ma:2014jla}. The situation improves if $P^z$ becomes much larger than the scale of the proton mass~\cite{Gamberg:2014zwa}.

In this paper, we present a method to reconstruct a PDF by combining information from its Mellin moments and from the corresponding quasi-PDF. As stated above, the PDF at intermediate $x$ is well reproduced by the quasi-PDF. At larger $x$, we use a parametric expression to fit some of the Mellin moments of the PDF itself. We require that the quasi-PDF and the parametric expression are the same at a certain matching point $x_0$, including the value of their first derivative.  Since lattice calculations of quasi-PDFs at sufficiently large $P^z$ are missing, we test our method by using the diquark spectator approximation for up and down valence distributions of both unpolarized and helicity PDFs. However, in the future  our procedure can be used to improve the calculation of PDFs entirely based on lattice QCD results. We also study the dependence of our results on the choice of the matching point $x_0$ and of the proton momentum $P^z$.

In the following, in Sec.~\ref{sec:qPDF} we describe the quasi-PDF in the diquark spectator approximation. In Sec.~\ref{sec:model}, we discuss our reconstruction procedure in detail. In Sec.~\ref{sec:out}, we show our results, and in Sec.~\ref{sec:end} we draw some conclusions.

%%%%%%%%%%%%%%%%%%%%%%%%%%%%%%%%%%%%%%%%%%%

\section{Quasi-PDFs in the spectator diquark approximation}
\label{sec:qPDF}

In this section, we recall the operator definitions of the leading-twist unpolarized distribution $f_1 (x)$ and helicity distribution $g_1 (x)$, and of the corresponding quasi-PDFs $\tilde{f}_1 (x, P^z)$ and $\tilde{g}_1 (x, P^z)$. Then, we list the analytic expressions of these distributions in the spectator diquark approximation. Throughout the paper, we will represent 4-vectors with both their Minkowski or light-cone components. In the latter case, we define the light-like vectors $n_\pm$ satisfying $n_\pm^2 = 0, \, n_+\cdot n_-=1,$ and we describe a generic 4-vector $a^\mu$ as $a = [a^-,a^+,\bm{a}_T] $, where $a^\pm = a\cdot n_\mp = (a^0 \pm a^z) / \sqrt{2}$.

%%%%%%%%%%%%%%

\subsection{Operator definitions of PDFs and quasi-PDFs}
\label{sec:f1&g1}

The operator definition of the leading-twist unpolarized distribution $f_1 (x)$ and helicity distribution $g_1 (x)$ is given by~\cite{RevModPhys.67.157}
\begin{align}
f_1 (x) = \int_{-\infty}^{\infty} &\frac{d\xi^-}{4\pi} \, e^{-i\xi^- k^+} \nonumber \\
&\times \left\langle P \left| \bar{\psi} (\xi^-) \gamma^{+} \, U_{n_-} [ \xi^-, 0] \, \psi (0) \, \right| P \right\rangle \, , \label{eq:f1}
\\[0.2cm]
g_1 (x) = \int_{-\infty}^{\infty} &\frac{d\xi^-}{4\pi} \, e^{-i\xi^- k^+} \nonumber \\
&\left\langle P S \left| \bar{\psi} (\xi^-) \gamma^{+} \gamma_5 \, U_{n_-} [ \xi^-, 0] \, \psi (0) \, \right| P S \right\rangle \, , \label{eq:g1}
\end{align}
where $P$ is the four-momentum of a nucleon with mass $M$, moving along the $\hat{z}$ direction, {\it i.e.}
\begin{equation}
P^\mu = \left[ \sqrt{\left(P^z\right)^2 + M^2} , \, \bm{0}_T, \, P^z \right] = \left[ P^- , \, P^+ , \, \bm{0}_T \right] \, ,
\label{eq:Pmu}
\end{equation}
$S$ is the longitudinal polarization of the nucleon with $S^2 = -1$ and $P \cdot S = 0$, and $x = k^+ / P^+$. The gauge link operator $U_{n_-}$ along the light-cone direction $n_-$ is given by
\begin{align}
U_{n_-} [ \xi^-, 0] &= {\cal P} \left[ \exp \left( -ig \int_{0}^{\xi^-} dw^- \, A^+ ( w^- ) \right) \, \right] \, ,
\label{eq:gaugelink}
\end{align}
and it connects the quark fields $\psi$ in the two different points $0$ and $(\xi^-, 0, \bm{0}_T)$ by all possible ordered paths followed by the gluon field $A$ with coupling $g$, thus  making the definitions in Eqs.~\eqref{eq:f1} and \eqref{eq:g1} gauge invariant.

The corresponding definitions of the quasi-PDFs $\tilde{f}_1 (x, P^z)$ and $\tilde{g}_1 (x, P^z)$ involve only spatial correlations along the $\hat{z}$ direction~\cite{Ji:2013dva}:
\begin{align}
\tilde{f}_1 (x, P^z) = \int_{-\infty}^{\infty} &\frac{d\xi^z}{4\pi} \, e^{i\xi^z k^z} \nonumber \\
&\times \left\langle P \left| \bar{\psi} (\xi^z) \gamma^z \, U_{z} [ \xi^z, 0] \, \psi (0) \, \right| P \right\rangle \, , \label{eq:qf1}
\\[0.2cm]
\tilde{g}_1 (x, P^z) = \int_{-\infty}^{\infty} &\frac{d\xi^z}{4\pi} \, e^{i\xi^z k^z} \nonumber \\
&\hspace{-0.5cm} \times \left\langle P S \left| \bar{\psi} (\xi^z) \gamma^z \gamma_5 \, U_{z} [ \xi^z, 0] \, \psi (0) \, \right| P S \right\rangle \, , \label{eq:qg1}
\end{align}
where now $x = k^z / P^z$ and the gauge link along the $\hat{z}$ direction takes the form
\begin{align}
U_{z} [ \xi^z, 0] &= {\cal P} \left[ \exp \left( -ig \int_{0}^{\xi^z} dw^z \, A^z ( w^z ) \right) \, \right] \, .
\label{eq:zgaugelink}
\end{align}

%%%%%%%%%%%%%%

\subsection{Unpolarized PDFs and quasi-PDFs in the spectator diquark model}
\label{sec:Diqf1}

The spectator diquark approximation consists of two basic steps~\cite{Jakob:1997wg}. First, we insert a completeness relation with intermediate states into the operator definition of PDFs, Eqs.~\eqref{eq:f1} and \eqref{eq:g1}. Then, we truncate the sum to a single on-shell spectator state of mass $M_X$ representing either a scalar diquark $(X=s)$ or an axial-vector diquark, which in turn can have isoscalar $(X=a)$ or isovector $(X=a')$ quantum numbers~\cite{Bacchetta:2008af}. The nucleon-quark-diquark interaction vertex can be dressed by a suitable form factor that can be chosen in different forms~\cite{Bacchetta:2008af}. Following Ref.~\cite{Gamberg:2014zwa}, we adopt the dipolar form
\begin{align}
{\cal I}_X (k^2) &= g_X \, \frac{k^2-m^2}{(k^2-\Lambda_X^2)^2} \, ,
\label{eq:ffdip}
\end{align}
where $m$ is the mass of a constituent quark with four-momentum $k$, $g_X$ and $\Lambda_X$ are appropriate coupling constants and cutoffs, respectively, to be considered as free parameters of the model together with the diquark mass $M_X$.

If we are sensitive also to the transverse component $\bm{k}_T$ of the parton momentum with respect to the direction of the nucleon momentum, the form factor can be conveniently rewritten as~\cite{Bacchetta:2008af}
\begin{align}
{\cal I}_X (k^2) &= g_X \, \frac{(k^2-m^2)\,(1-x)^2}{\left( \bm{k}_T^2 + L_X^2 (\Lambda_X^2) \right)^2} \, ,
\label{eq:ffdip2}
\end{align}
where the function $L_X$ is given by
\begin{align}
L_X^2 (\Lambda_X^2) &= x M_X^2 + (1-x) \Lambda_X^2 - x (1-x) M^2 \, ,
\label{eq:L}
\end{align}
and it is useful to define the off-shell condition for the quark~\cite{Bacchetta:2008af}:
\begin{align}
k^2 - m^2 &= -\frac{\bm{k}_T^2 + L_X^2(m^2)}{1-x}  \, .
\label{eq:offshell}
\end{align}

In the lowest order, the expressions of the unpolarized distributions for the scalar $(f_1^s)$ and axial-vector $(f_1^a)$ diquarks become~\cite{Bacchetta:2008af,Gamberg:2014zwa}
\begin{align}
f_1^s (x, \bm{k}_T^2) &= \frac{g_s^2}{(2\pi)^3}\, \frac{[(m+xM)^2 + \bm{k}_T^2] \, (1-x)^3}
                                                                                         {2 \, [\bm{k}_T^2 + L_s^2 (\Lambda_s^2)]^4}  \, ,
\label{eq:f1s}
\\[0.2cm]
f_1^a (x, \bm{k}_T^2) &= \frac{g_a^2}{(2\pi)^3}\nonumber \\
&\times  \frac{[ \bm{k}_T^2 \, (1+x^2) + (m+xM)^2 \, (1-x)^2] \, (1-x)}
                                                                                          {2 \, [\bm{k}_T^2 + L_a^2 (\Lambda_a^2)]^4} \, .
\label{eq:f1a}
\end{align}

The corresponding quasi-PDFs $\tilde{f}_1^s (x, \bm{k}_T^2, P^z)$ and $\tilde{f}_1^a (x, \bm{k}_T^2, P^z)$ read~\cite{Gamberg:2014zwa}
\begin{align}
\tilde{f}_1^s (x, \bm{k}_T^2, P^z) &= [{\cal I}_s (k^2)]^2  \, \frac{{\cal F}_s}{{\cal D}_s}  \, ,
\label{eq:qf1s}
\\[0.2cm]
\tilde{f}_1^a (x, \bm{k}_T^2, P^z) &= [{\cal I}_a (k^2)]^2 \, \Big[ (M_a^2 + (1-x)^2\, \left(P^z\right)^2) \, {\cal D}_a \Big]^{-1} \nonumber \\[0.1cm]
&\hspace{-1.5cm} \times \Big[ (M_a^2 + (1-x)^2\, \left(P^z\right)^2) \, ({\cal F}_a - 2x M^2) \nonumber \\
&\hspace{-1cm} - 2x (1-x)^2 \left( P^z \right)^4 (1- \rho_a^2 \delta^2) - 2x M_a^2 \left( P^z \right)^2 \Big] ,
\label{eq:qf1a}
\end{align}
where
\begin{align}
{\cal F}_X &= (2x -1 ) M^2 + 2x M m - M_X^2 + m^2 \nonumber \\
&\quad - 2 (1-x)^2 \, \left(P^z\right)^2 \, (1 - \rho_X \, \delta ) \, , \\
{\cal D}_X &= \Big[ 2 (1-x) \, \left(P^z\right)^2 \, (1 - \rho_X \, \delta ) + M^2 + M_X^2 - m^2 \Big]^2 \nonumber \\
&\quad \times 2 (1-x) \, \rho_X \, ,
\label{eq:service}
\end{align}
and
\begin{align}
\rho_X &= \sqrt{1 + \frac{\bm{k}_T^2 + M_X^2}{(1-x)^2\, \left(P^z\right)^2}} \, , \\
\delta &= \sqrt{1 + \frac{M^2}{\left(P^z\right)^2}} \, .
\label{eq:service2}
\end{align}

It is easy to verify that in the limit of large $\left(P^z\right)^2 \gg M^2, \, M_X^2$ we have~\cite{Gamberg:2014zwa}
\begin{align}
\lim_{P^z \to \infty} \tilde{f}_1^X (x, \bm{k}_T^2, P^z) &= f_1^X (x, \bm{k}_T^2) \, .
\label{eq:qPDFlimPDF}
\end{align}
However, this result holds only if $x$ is not very large. Otherwise, the term $(1-x)^2\, \left( P^z \right)^2$ breaks down the large-$P^z$ expansion that leads to Eq.~\eqref{eq:qPDFlimPDF}. Therefore, we have to expect that the quasi-PDFs are not a good approximation to standard PDFs in the large $x$ region, unless $P^z$ is boosted to very large values~\cite{Gamberg:2014zwa}.

By integrating Eqs.~\eqref{eq:f1s} and \eqref{eq:f1a} upon the parton transverse momentum, we get the diquark scalar and axial-vector components of the unpolarized collinear PDF~\cite{Bacchetta:2008af}:
\begin{align}
f_1^s (x) &= \frac{g_s^2}{(2\pi)^2}\,
\frac{[2\,(m+xM)^2 + L_s^2 (\Lambda_s^2) ] \, (1-x)^3}{24 \, L_s^6 (\Lambda_s^2)}  \, ,
\label{eq:f1sPDF}
\\
f_1^a (x) &= \frac{g_a^2}{(2\pi)^2}\nonumber \\
&\hspace{-0.5cm} \times \frac{[2\,(m+xM)^2 \, (1-x)^2 + (1+x^2) \, L_a^2 (\Lambda_a^2) ] \, (1-x)}
        {24 \, L_a^6 (\Lambda_a^2)}  \, .
\label{eq:f1aPDF}
\end{align}
The corresponding expressions for the quasi-PDFs $\tilde{f}_1^s (x, P^z)$ and $\tilde{f}_1^a (x, P^z)$ are very lengthy and are shown in Appendix~\ref{sec:Af1}. It is straightforward to verify that in the limit $P^z \to \infty$ they recover the corresponding PDFs of Eqs.~\eqref{eq:f1sPDF} and~\eqref{eq:f1aPDF}.

%%%%%%%%%%%%%%

\subsection{Helicity PDFs and quasi-PDFs in the spectator diquark model}
\label{sec:Diqg1}

Following the same steps of the previous section, the helicity distributions for the scalar $(g_1^s)$ and axial-vector $(g_1^a)$ diquarks are~\cite{Bacchetta:2008af,Gamberg:2014zwa}
\begin{align}
g_1^s (x, \bm{k}_T^2) &= \frac{g_s^2}{(2\pi)^3}\, \frac{[(m+xM)^2 - \bm{k}_T^2] \, (1-x)^3}
                                                                                         {2 \, [\bm{k}_T^2 + L_s^2 (\Lambda_s^2)]^4}  \, ,
\label{eq:g1s}
\\[0.2cm]
g_1^a (x, \bm{k}_T^2) &= \frac{g_a^2}{(2\pi)^3}\nonumber \\
&\times \frac{[ \bm{k}_T^2 \, (1+x^2) - (m+xM)^2 \, (1-x)^2] \, (1-x)}
                      {2 \, [\bm{k}_T^2 + L_a^2 (\Lambda_a^2)]^4} \, .
\label{eq:g1a}
\end{align}

The corresponding quasi-PDFs $\tilde{g}_1^s (x, \bm{k}_T^2, P^z)$ and $\tilde{g}_1^a (x, \bm{k}_T^2, P^z)$ read~\cite{Gamberg:2014zwa}
\begin{align}
\tilde{g}_1^s (x, \bm{k}_T^2, P^z) &= [{\cal I}_s (k^2)]^2  \, \frac{{\cal G}_s}{{\cal D}_s}  \, ,
\label{eq:qg1s}
\\
\tilde{g}_1^a (x, \bm{k}_T^2, P^z) &= [{\cal I}_a (k^2)]^2 \Big[ M (M_a^2 + (1-x)^2 \left(P^z \right)^2) \, {\cal D}_a \Big]^{-1} \nonumber \\
&\hspace{-1.5cm} \times \Big[ [M_a^2 + (1-x)^2 \left( P^z \right)^2] \, (2M^2 \delta m - {\cal G}_a) \nonumber \\
&\hspace{-1cm} + 2 (1-x)^2 \left( P^z \right)^4 \delta [ \rho_a^2 \, (xM + m\, (1 - \delta^2)) - xM ] \nonumber \\
&\hspace{-1cm} - 2x M_a^2 \left( P^z \right)^2 \Big] \, ,
\label{eq:qg1a}
\end{align}
where
\begin{align}
{\cal G}_X &= 2\, (1-x) \, \rho_X \, \left( P^z \right)^2\, [ (x - \delta^2)\, M + (1-\delta^2)\, m] \nonumber \\
&\quad + \delta M \, [ (M+m)^2 + M_X^2 + 2 (1-x)^2 \left( P^z \right)^2 ] \, .
\label{eq:service3}
\end{align}

The integrated helicity PDFs are~\cite{Bacchetta:2008af}
\begin{align}
g_1^s (x) &= \frac{g_s^2}{(2\pi)^2}\,
\frac{[2\,(m+xM)^2 - L_s^2 (\Lambda_s^2) ] \, (1-x)^3}{24 \, L_s^6 (\Lambda_s^2)}  \, ,
\label{eq:g1sPDF}
\\
g_1^a (x) &= \frac{g_a^2}{(2\pi)^2}\nonumber \\
&\hspace{-0.5cm} \times \frac{[2\,(m+xM)^2 \, (1-x)^2 - (1+x^2) \, L_a^2 (\Lambda_a^2) ] \, (1-x)}
                       {24 \, L_a^6 (\Lambda_a^2)}  \, .
\label{eq:g1aPDF}
\end{align}
Again, the expressions for the quasi-helicities $\tilde{g}_1^s (x, P^z)$ and $\tilde{g}_1^a (x, P^z)$ are reported in Appendix~\ref{sec:Ag1}. It is straightforward to show that~\cite{Gamberg:2014zwa} 
\begin{align}
\lim_{P^z \to \infty} \tilde{g}_1^X (x, \bm{k}_T^2, P^z) &= g_1^X (x, \bm{k}_T^2) \, , \\
\lim_{P^z \to \infty} \tilde{g}_1^X (x, P^z) &= g_1^X (x) \, .
\label{eq:qglimg}
\end{align}

%%%%%%%%%%%%%%

\subsection{Plots of unpolarized and helicity quasi-PDFs}
\label{sec:qPDFplot}

Following Ref.~\cite{Bacchetta:2008af}, we combine the above results for the scalar and axial-vector diquark components to give the up and down unpolarized and helicity PDFs. In order to keep simple the probabilistic interpretation of the results, it is convenient to use normalized versions of $f_1^X$ and $g_1^X$. Therefore, for example we use $||f_1^X|| = (N_X^2 / g_X^2)\, f_1^X$ where the normalization $N_X$ is determined by requiring~\cite{Bacchetta:2008af} 
\begin{align}
\pi \int_0^1 dx \int_0^{\infty} d\bm{k}_T^2 \, ||f_1^X|| (x,\bm{k}_T^2) &= 1 \, .
\label{eq:norm}
\end{align}
The flavor components of PDFs are given by 
\begin{align}
f_1^u &= c_s^2\, ||f_1^s|| + c_a^2\, ||f_1^a|| \, ,
\label{eq:uX} \\
f_1^d &= c_a^{\prime 2}\, ||f_1^{a'}||  \; ,
\label{eq:dX}
\end{align}
and similarly for the helicity PDF $g_1$. Hence, the up quark receives contributions from both the scalar-isoscalar $(s)$ and from the axial-vector-isoscalar $(a)$ diquark components, while the down quark is completely determined by the axial-vector-isovector component $(a')$. The coefficients $c_X$ of the linear combination are free parameters of the model. Together with the diquark masses $M_X$ and cutoffs $\Lambda_X$, they were fixed in Ref.~\cite{Bacchetta:2008af} by fitting the ZEUS parametrization of $f_1^u (x)$ and $f_1^d(x)$ at $Q_0^2= 0.3$ GeV$^2$ (ZEUS2002)~\cite{Chekanov:2002pv}, and the leading-order parametrization of $g_1^u (x)$ and $g_1^d(x)$ at $Q_0^2 = 0.26$ GeV$^2$ from Ref.~\cite{Gluck:2000dy} (GRSV2000). The $c_X$ play the role of "effective couplings" and are related to the original coupling constants of the model by $c_X^2 \, N^2_X = g_X^2$, with $X = s, a, a'$~\cite{Bacchetta:2008af}.

%%%%%% Fig. 1 - quasi-PDF vs. PDF
\begin{figure}
\begin{center}
\includegraphics[width=0.5\textwidth]{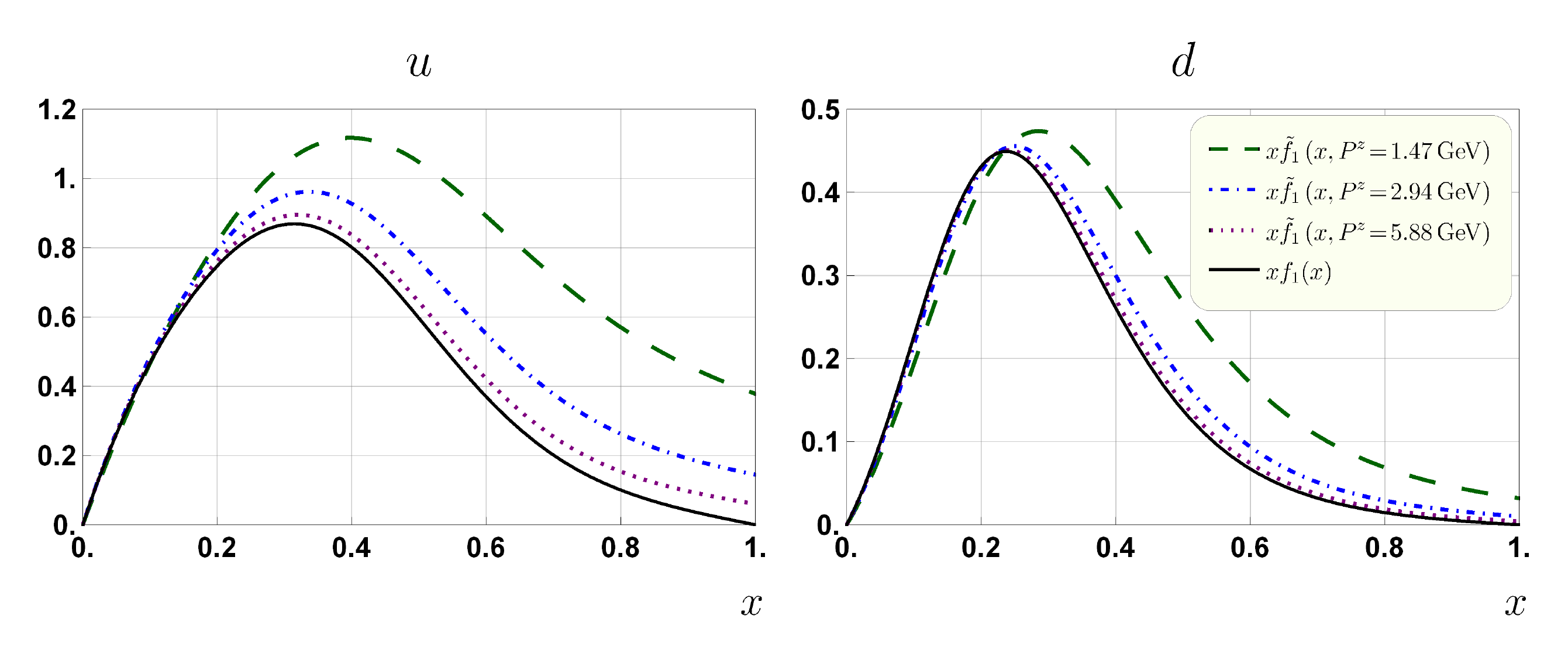} \\[0.3cm]
\includegraphics[width=0.5\textwidth]{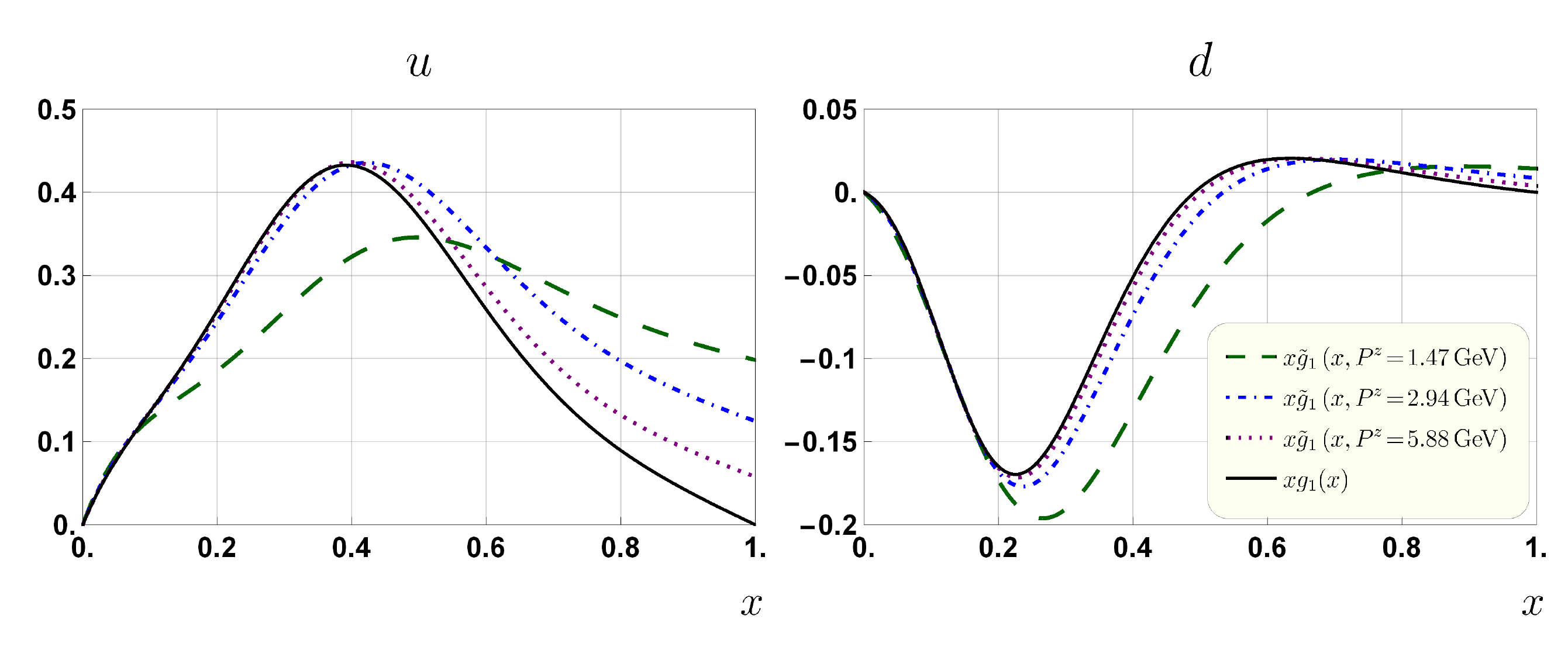}
\end{center}
\vspace{-0.5cm}
\caption{\label{fig:cmpr_LC_QS} Comparison between PDFs and quasi-PDFs. Upper panels for $f_1$ versus 
$\tilde{f}_1$, lower panels for $g_1$ versus $\tilde{g}_1$. Left panels for the up quark, right panels for the down
 quark. In all panels, solid lines for the standard PDFs calculated in the diquark spectator model of 
Ref.~\cite{Bacchetta:2008af}. Dashed lines for the quasi-PDFs at $P^z = 1.47$ GeV, dot-dashed lines at $P^z = 2.94$ GeV, dotted lines at $P^z = 5.88$ GeV.}
\end{figure}
%%%%%%%%%%%%%

Using the parameter values listed in Tab.~I of Ref.~\cite{Bacchetta:2008af}, we can calculate both up and down components of the unpolarized PDF $f_1$ and the helicity PDF $g_1$, as well as the corresponding quasi-PDFs $\tilde{f}_1$ and $\tilde{g}_1$ at various values of $P^z$. The comparison between quasi-PDFs and PDFs is shown in Fig.~\ref{fig:cmpr_LC_QS}. The upper panels show the results for $f_1$ versus $\tilde{f}_1$, the lower panels for $g_1$ versus $\tilde{g}_1$. The left panels display the results for the up quark, the right panels for the down quark. In all panels, the solid lines refer to the standard PDFs calculated in the diquark spectator model of Ref.~\cite{Bacchetta:2008af}. The dashed lines are the quasi-PDFs computed at $P^z = 1.47$ GeV. The dot-dashed lines refer to $P^z = 2.94$ GeV, the dotted lines to $P^z = 5.88$ GeV. It is evident that the quasi-PDFs better approximate the corresponding PDFs as $P^z$ increases.

%%%%%%%%%%%%%%%%%%%%%%%%%%%%%%%%%%%%%%%%%%%

\section{The reconstruction procedure}
\label{sec:model}

From Fig.~\ref{fig:cmpr_LC_QS}, it is already evident that the quasi-PDFs are always a good approximation to PDFs for $x \lesssim 0.2$, and that there is a discrepancy at larger $x$ which becomes smaller for increasing $P^z$. But large values of $P^z$ are currently beyond the reach of lattice computations of quasi-PDFs. In the following, we describe a procedure to reconstruct the PDF at large $x$ by using information on the corresponding quasi-PDF and on the Mellin moments of the PDF itself.

We choose a point $x_0$, denoted matching point, that divides the support of PDFs in two regions: the lower-$x$ region $0 \leq x \leq x_0$, and the higher-$x$ region $x_0 < x \leq 1$. At some factorization scale $\mu$ (that will be mostly understood for simplicity), we assume that the PDF $q(x, \mu)$ for flavor $q$ is well approximated by the quasi-PDF $\tilde{q} (x, \mu, P^z)$ in the lower-$x$ region. In the higher-$x$ region, we replace the quasi-PDF with the parametric expression
\begin{align}
\hat{q} \left( x, \left\{ p_i \right\} \right) &= x^{p_1} \left( 1 - x \right)^{p_2} \left( 1+ p_3 \, x^{1/2} + p_4 \, x + p_5 \, x^{3/2} \right) \, ,
\label{eq:param}
\end{align}
where $p_2 > 0$ because the standard PDF vanishes when $x \geq 1$.

We require that the lower- and higher-$x$ regions are smoothly connected at $x = x_0$ by imposing that the quasi-PDF and the parametric expression coincide with their value and first derivative, namely
\begin{align}
\tilde{q} \left( x_{0}, P^z \right) &= \hat{q} \left( x_{0}; \left\{ p_i \right\} \right) \, , \\
\frac{d}{dx} \tilde{q} \left( x, P^z \right) \Big|_{x=x_{0}} &= \frac{d}{dx} \hat{q} \left( x; \left\{ p_i \right\} \right) \Big|_{x=x_{0}} \, .
\label{eq:match}
\end{align}
Hence, we can eliminate two free parameters. We choose to represent $p_3$ and $p_4$ as analytic functions of $p_1, \, p_2,$ and $p_5$. These latter free parameters are further determined by minimizing the weighted square distance $\chi^2$ between the $n=2,\, 3, \, 4$ (truncated) Mellin moments of the quasi-PDF $\tilde{q}$ and the parametric function $\hat{q}$ with respect to the standard PDF $q$,
\begin{align}
\chi^{2} \left( \left\{ p_1, p_2, p_5 \right\} \right) &= \sum_{n=2}^4 \frac{\big[ \hat{q}^{n} \left( \left\{ p_1, p_2, p_5 \right\} \right) +\tilde{q}^{\, n} ( P^z ) - q^{n} \big]^2}{\big[ \tilde{q}^n \left( P^z \right) - q^n \big]^2} \, ,
\label{eq:chi2}
\end{align}
where
\begin{align}
q^n &= \int_0^1 dx \, x^{n-1} \, q(x) \, ,  \nonumber \\
\tilde{q}^n (P^z) &= \int_0^{x_0} dx \, x^{n-1} \, \tilde{q} (x, P^z) \, , \nonumber \\
\hat{q}^n  \left( \left\{ p_1, p_2, p_5 \right\} \right) &= \int_{x_0}^1 dx \, x^{n-1} \,  \hat{q}  \left( \left\{ p_1, p_2, p_5 \right\} \right) \, ,
\label{eq:Mellin}
\end{align}
and in all formulae the dependence on the factorization scale $\mu$ is understood. The weights $ \left[ \tilde{q}^n \left( P^z \right) - q^n \right]^{-2}$ are added to balance the importance of each moment. 

When we try to reconstruct the unpolarized PDF $\hat{q} \left( x; \left\{ p_i \right\} \right) \equiv \hat{f}_1^q (x; \left\{ p_i \right\} )$, we further impose that in the higher-$x$ region
\begin{align}
\frac{d \hat{f}_1^q \left(x; \left\{ p_i \right\} \right)}{dx} &< 0 \, , \quad
\frac{d^{2} \hat{f}_1^q \left( x; \left\{ p_i \right\} \right)}{dx^2} > 0 \, .
\label{eq:f1concavity}
\end{align}
This constraint reflects the concavity of the unpolarized PDF at large $x$, namely the fact that for increasing $x$ the unpolarized PDF always decreases while its first derivative increases. In practice, the constraint in Eq.~\eqref{eq:f1concavity} is implemented by sampling ten points $x_i$ uniformly distributed in the range $x_0 < x_i < 1$, with $i = 1,..,10$.

The choice of the matching point $x_0$ is arbitrary. The only guidance is that in the lower-$x$ region $0 \leq x \leq x_0$ the expressions $x^{n-1}\, \tilde{q} (x, P^z)$ with $n = 2,..,4$, involving the quasi-PDF, are close enough to the corresponding ones for the standard PDF $q (x)$ such that the truncated Mellin moments are almost identical. It turns out, also by inspecting Fig.~\ref{fig:cmpr_LC_QS}, that $0.2 \leq x_0 \leq 0.3$ is a convenient range. In the following, we will run our procedure both for $x_0 = 0.2$ and $0.3$. This choice is somewhat more conservative with respect to the one of Ref.~\cite{Gamberg:2014zwa}, where the authors claim that the quasi-PDFs are a good approximation of standard PDFs up to $x \simeq 0.4$. We have tried the option $x_0 = 0.4$, but in this case our reconstruction procedure fails.

When minimizing the function $\chi^{2} \left( \left\{ p_1, p_2, p_5 \right\} \right)$ in Eq.~\eqref{eq:chi2}, we consider only the $n=2,\, 3, \, 4$ (truncated) Mellin moments. The first Mellin moment is excluded because lattice calculations of the quasi-PDF are not reliable at small $x$. In fact, the largest nucleon momentum that can be generated on lattice is of the order $P_{\mathrm{max}} = a^{-1}$, where $a$ is the lattice spacing. The lowest momentum is $P_{\mathrm{min}} = \left( L a\right)^{-1}$, where $L$ is the number of lattice spacing in the $\hat{z}$ direction. Therefore, the smallest momentum fraction that can be simulated on lattice is $x_{\mathrm{min}} = P_{\mathrm{min}} / P_{\mathrm{max}} = L^{-1}$. The  largest lattice used in current calculations of quasi-PDFs is $32^3 \times 64$~\cite{Alexandrou:2015rja} so that the corresponding smallest momentum fraction is $x_{\mathrm{min}} = 1/32$. The contribution from the region $0 \leq x \leq x_{\mathrm{min}}$ to the various truncated Mellin moments can be estimated by taking its ratio to the full Mellin moments,
\begin{align}
\Delta^{(n)}_q (P^z) &= \frac{\int_0^{x_{\mathrm{min}}} dx \, x^{n-1} \tilde{q} (x, P^z)}{\int_0^1 dx \, x^{n-1} \tilde{q} (x, P^z)} \, .
\label{eq:truncfull}
\end{align}

%\vspace{-0.1cm}
%%%%%% Fig. 2 - truncated vs. full Mellin moments
\begin{figure}[h]
\begin{center}
\includegraphics[width=0.38\textwidth]{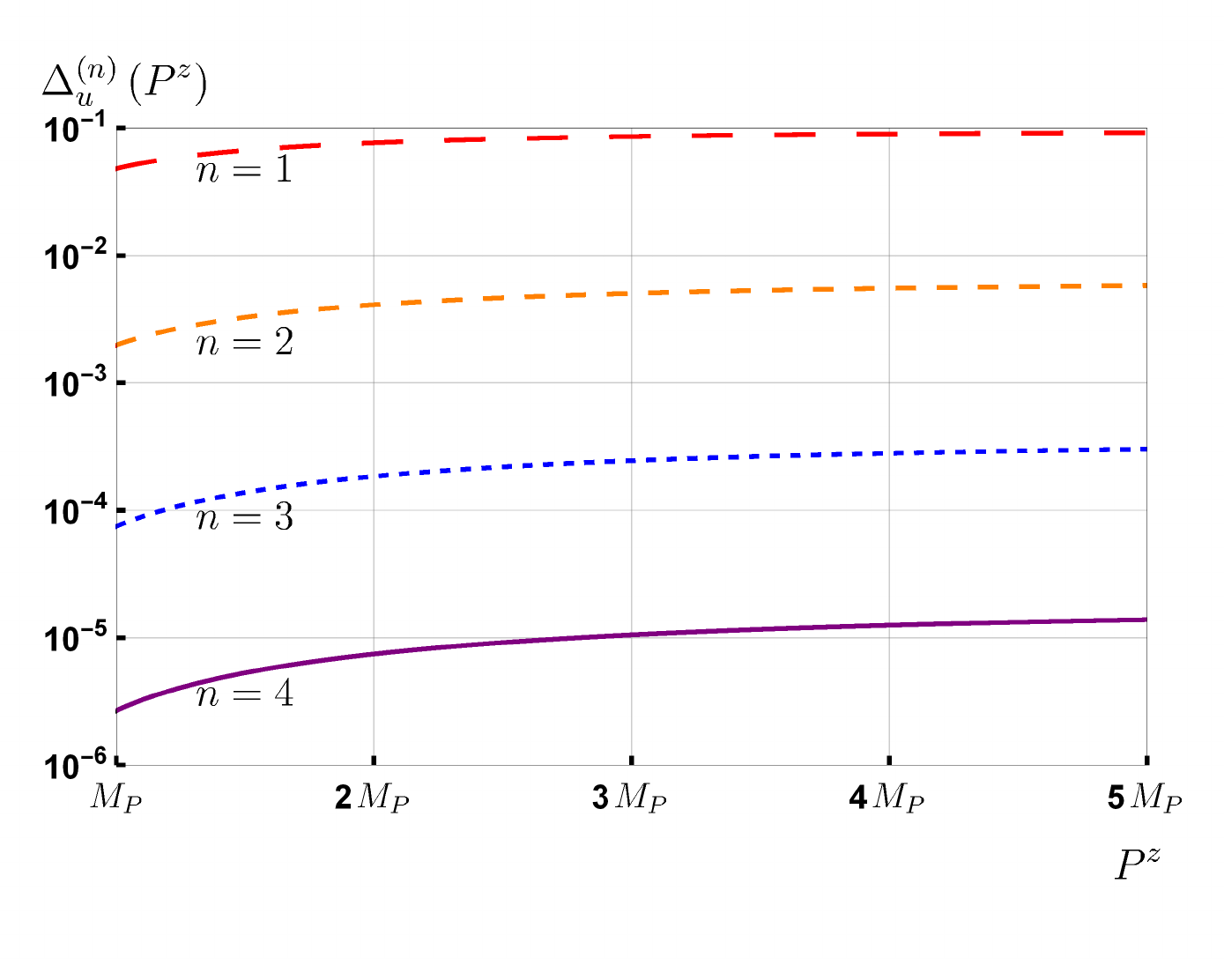}  \\[-0.8cm]
\includegraphics[width=0.38\textwidth]{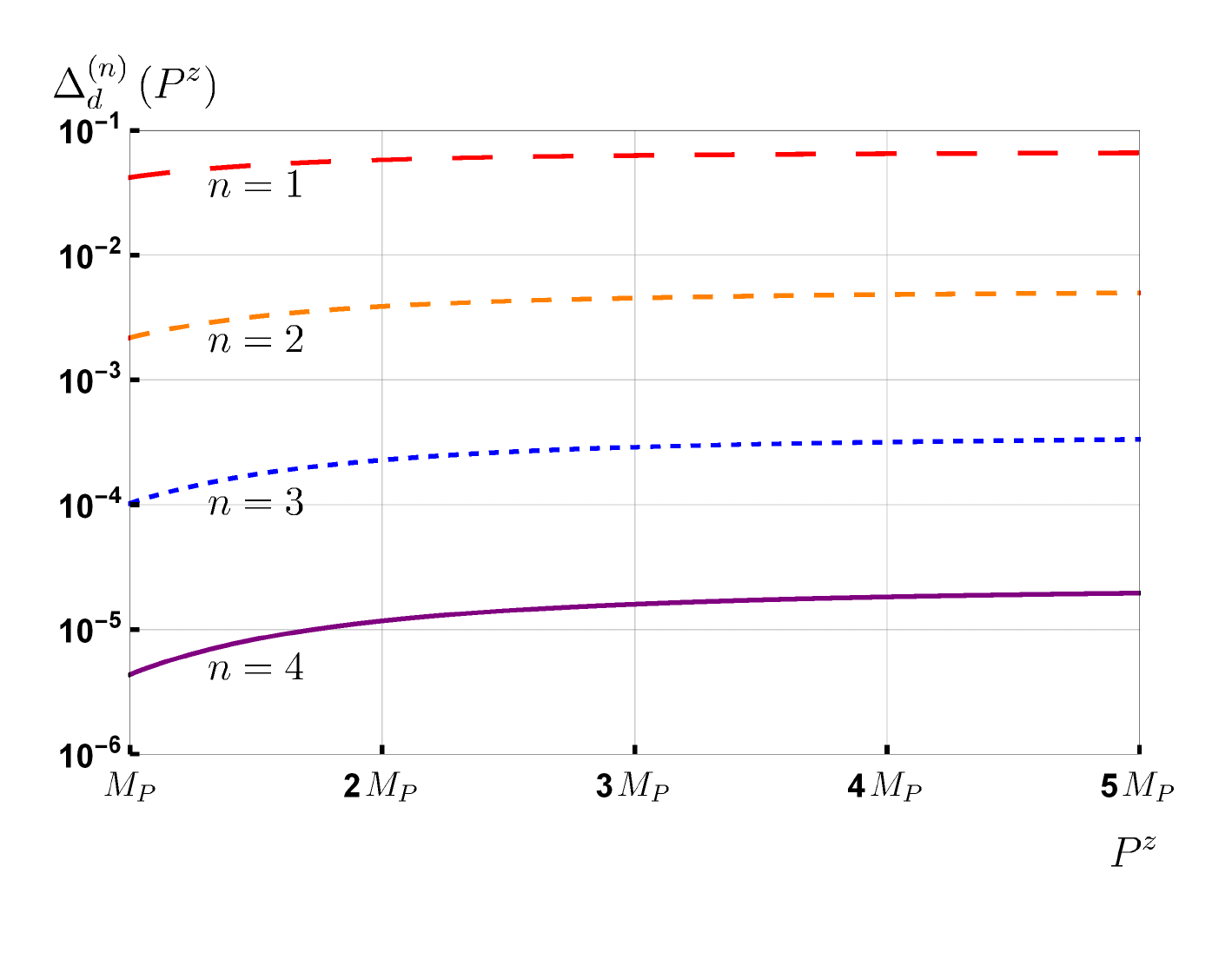}
\end{center}
%\vspace{-1.2cm}
\caption{\label{fig:Dlt_n} The $\Delta^{(n)}_q (P^z)$ of Eq.~\eqref{eq:truncfull} as a function of the nucleon longitudinal momentum $P^z$, in multiples of the proton mass $M_P$. Upper panel for the up quark, lower panel for the down quark. From top to bottom, long-dashed line for the $n = 1$ Mellin moment, medium-dashed for the $n = 2$, short-dashed for the $n = 3$, solid for the $n = 4$. The unpolarized quasi-PDF $\tilde{q} (x, P^z) \equiv \tilde{f}_1^q (x, P^z)$ is evaluated at the diquark spectator model scale $\mu^2 = Q_0^2 = 0.3$ GeV$^2$.}
\end{figure}
%%%%%%%%%%%%%
%\vspace{-0.5cm}

In Fig.~\ref{fig:Dlt_n}, the numerical results of $\Delta^{(n)}_q (P^z)$ for the diquark spectator model calculation of the unpolarized quasi-PDF $\tilde{q} (x, P^z) \equiv \tilde{f}_1^q (x, P^z)$ are presented as functions of $P^z$, which can range over several multiples of the proton mass $M_P$. The upper panel refers to the up quark $q = u$, the lower panel to the down quark $q = d$. From top to bottom, the long-dashed line refers to the $n = 1$ moment, the medium-dashed to $n = 2$, the short-dashed to $n = 3$, and the solid to $n = 4$. The quasi-PDF is evaluated at the natural model scale $\mu^2 = Q_0^2 = 0.3$ GeV$^2$. It is evident that the first truncated Mellin moment (top long-dashed curve) is as large as 10\% of the corresponding full moment, while the other truncated higher moments contribute to much less than 1\% of the corresponding full moments. Hence, the uncertainty coming from the $0 \leq x \leq x_{\mathrm{min}}$ region is negligible if we use the $n = 2,..,4$ Mellin moments of quasi-PDFs in the minimization formula~\eqref{eq:chi2} that fixes the parameters $\left\{ p_1, \, p_2, \, p_5 \right\}$.

%%%%%%%%%%%%%%%%%%%%%%%%%%%%%%%%%%%%%%%%%%%%%%%%

\section{Results of the reconstruction procedure}
\label{sec:out}

In this section, we present the results of our reconstruction procedure for both unpolarized and helicity PDFs. We compare the results for the standard PDF $q (x)$, computed in the diquark spectator model (as described in Sec.~\ref{sec:Diqf1} for $q (x) \equiv f_1^q (x)$ and in Sec.~\ref{sec:Diqg1} for $q (x) \equiv g_1^q (x)$, respectively) for $q = u, d$ at the model scale $Q_0^2$, for the corresponding quasi-PDF $\tilde{q} (x, P^z)$ (similarly, described in Sec.~\ref{sec:Diqf1} for $\tilde{q} (x, P^z) \equiv \tilde{f}_1^q (x, P^z)$ and in Sec.~\ref{sec:Diqg1} for $\tilde{q} (x, P^z) \equiv \tilde{g}_1^q (x, P^z)$), and for our reconstructed PDF $\overset{\circ}{q} (x, P^z)$, defined in Sec.~\ref{sec:model} as
\begin{align}
\overset{\circ}{q} (x, P^z) &= \begin{cases}
\tilde{q} (x, P^z) & 0 \leq x \leq x_0 \\
\hat{q} \left( x; \left\{ p_i \right\} \right) & x_{0} < x \leq 1
\end{cases} \, ,
\label{eq:QPDFdef}
\end{align}
where the parametric expression $\hat{q} \left( x; \left\{ p_i \right\} \right)$ is defined in Eq.~\eqref{eq:param}, subject to the constraints of Eqs.~\eqref{eq:match} and \eqref{eq:f1concavity}. In all cases, we consider the PDFs multiplied by the fractional momentum $x$. The matching point is fixed to $x_0 = 0.2$ or $0.3$. The reconstructed PDF $\overset{\circ}{q} (x, P^z)$ depends on  three parameters that can be fixed by minimizing the $\chi^2$ function defined in Eq.~\eqref{eq:chi2}.

%%%%%% Fig. 3 -reconstruction at x0=0.2, Pz=1.47 GeV
\begin{figure}[h]
\begin{center}
\includegraphics[width=0.46\textwidth]{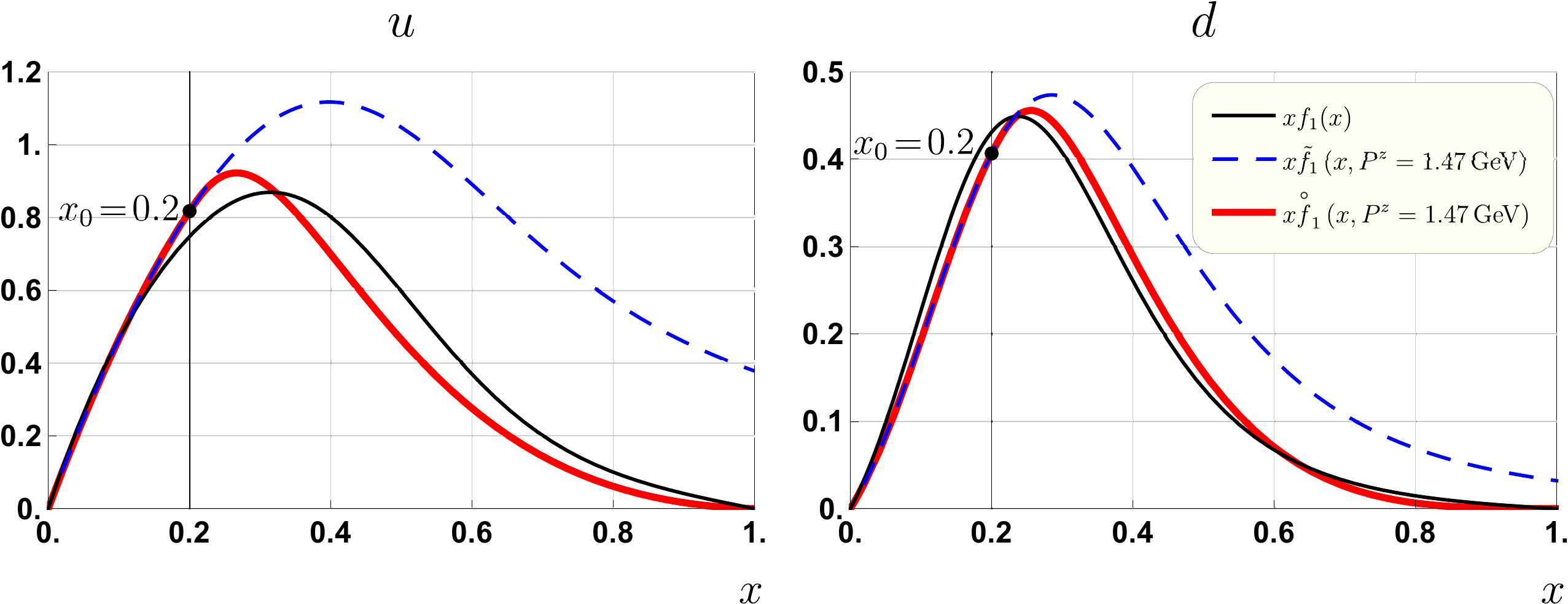} \\
\includegraphics[width=0.46\textwidth]{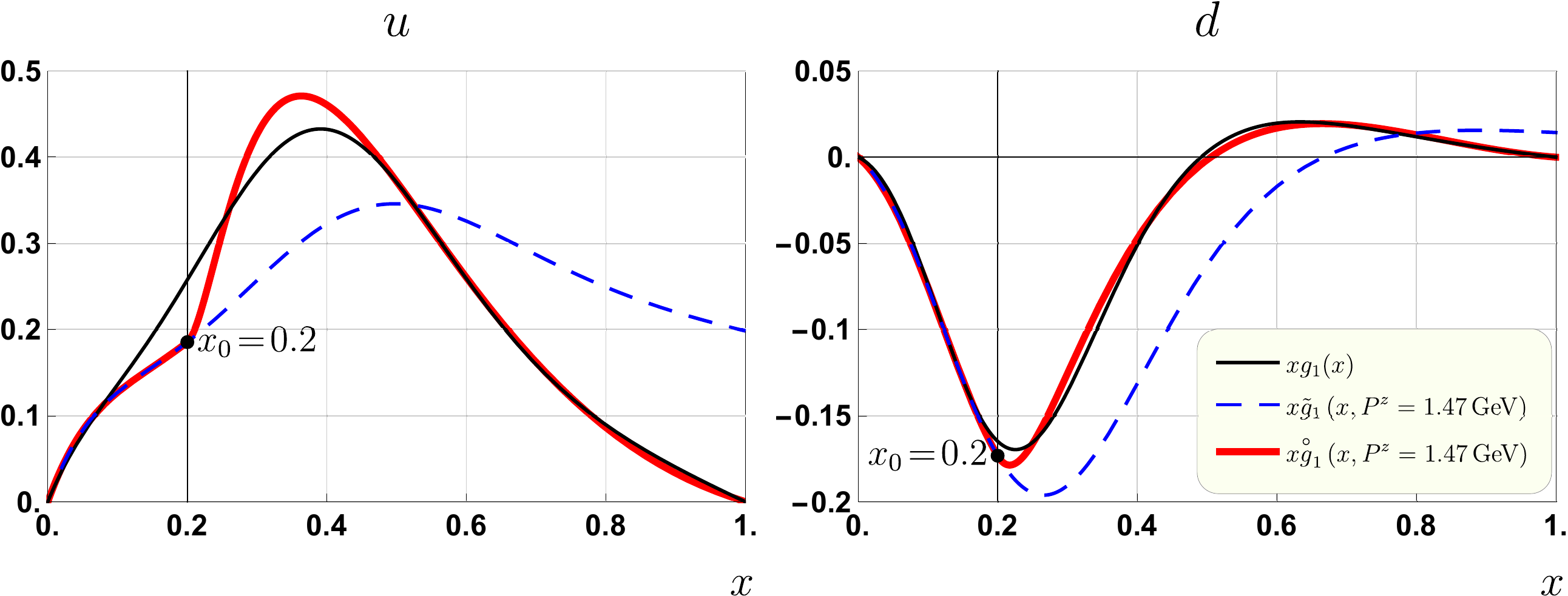}
\end{center}
%\vspace{-0.5cm}
\caption{\label{fig:cmprsn_PDFs_x02Pz147} Comparison among the standard PDF $x q(x)$ (black solid line), the quasi-PDF $x \tilde{q} (x, P^z)$ (dashed line), and the reconstructed PDF $x \overset{\circ}{q} (x, P^z)$ (lighter solid line) at $x_0 = 0.2$ for $P^z = 1.47$ GeV. Upper panels for $q (x) \equiv f_1^q (x)$, lower panels for $q (x) \equiv g_1^q (x)$. Left panels for $q = u$, right panels for $q = d$.}
\end{figure}
%%%%%%%%%%%%%
%\vspace{-0.8cm}

In Fig.~\ref{fig:cmprsn_PDFs_x02Pz147}, the comparison is shown at $x_0 = 0.2$ for $P^z = 1.47$ GeV. The upper panels refer to the unpolarized PDF, the lower panels to the helicity PDF; the left panels show the results for the up quark, the right panels for the down quark. The standard PDFs are represented by black solid lines, the quasi-PDFs by dashed (blue) lines, the reconstructed PDFs by lighter (red) solid lines. The quasi-PDFs $x \tilde{q} (x)$ are a reliable reproduction of the PDFs $x q (x)$ only for $x \leq x_0$: at higher $x$, they largely deviate and do not show the correct asymptotic behaviour for $x \to 1$. Nevertheless, our parametric expressions $x \hat{q} (x)$ follow quite closely the PDFs $x q (x)$ at very large $x$. Though, some conspicuous oscillations around $x q(x)$ appear at intermediate $x \gtrsim x_0$, in particular for the up quark, suggesting that the overall agreement is not optimal.

\begin{figure}[h]
\begin{center}
\includegraphics[width=0.47\textwidth]{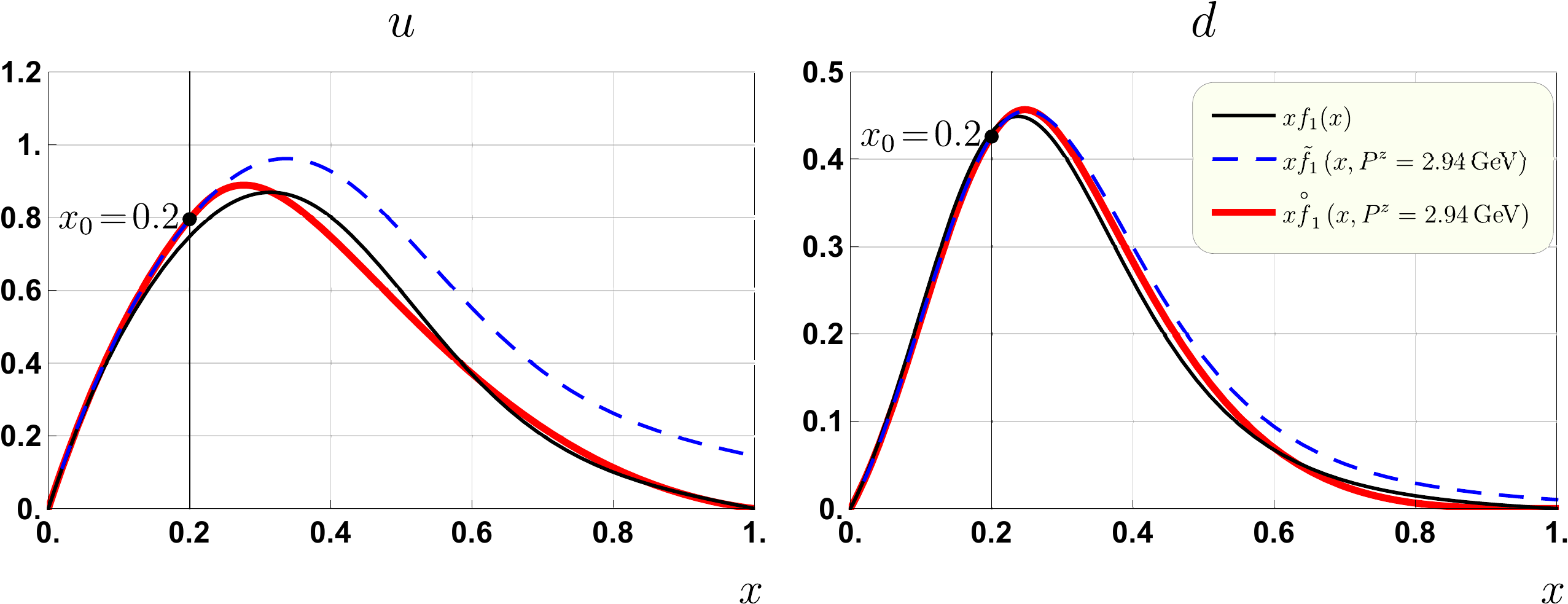}  \\
\includegraphics[width=0.47\textwidth]{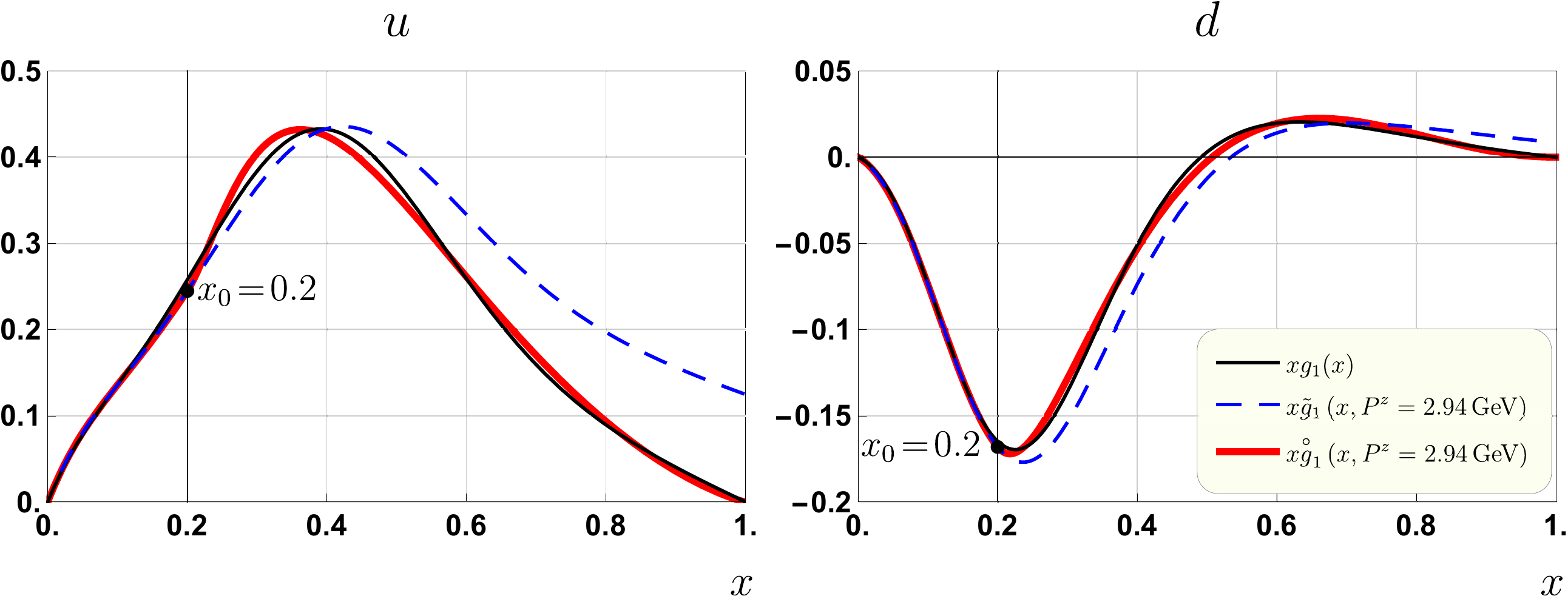}
\end{center}
%\vspace{-0.5cm}
\caption{\label{fig:cmprsn_PDFs_x02Pz294} Same notation and conventions as in the previous figure but for $P^z = 2.94$ GeV.}
\end{figure}

In Fig.~\ref{fig:cmprsn_PDFs_x02Pz294}, the same situation is reconsidered for $P^z = 2.94$ GeV. It is evident that increasing $P^z$ improves our reconstruction procedure because the quasi-PDF is already much closer to the standard PDF over a significant range of $x$ values. The reconstructed PDF $x \overset{\circ}{q} (x, P^z)$ looks like a close approximation to the standard PDF $x q (x)$ over the entire range $0 \leq x \leq 1$ for both unpolarized and helicity PDFs, with some minor oscillations around $x q(x)$ in the unpolarized up-quark channel at $x \gtrsim x_0$.

%%%%%%%%%
\begin{widetext}
\begin{center}

%%%%%%% Tab. I - parameter and distance values for x0=0.2

\begin{table}[!hbt]
\begin{tabular}{|c|c|c|c|c|c|c|c|c|}
\hline
$\begin{aligned}
x_0 &= 0.2\\
P^z &= 1.47 \; \mathrm{GeV}
\end{aligned}$
& $p_1$ & $p_2$ & $p_3$ & $p_4$ & $p_5$ & $\chi^{2}$ & $\overset{\circ}{r}$ & $\tilde{r}$ \tabularnewline
\hline
$f_1^u$ & -3.1067 & 1.4196 & -6.0771 & 11.543 & -6.1836 & 0.08493 & 4.0408 $\times 10^{-3}$ & 0.059932 \tabularnewline
\hline
$f_1^d$ & -3.0189 & 2.8007 & -5.9664 & 11.096 & -5.8289 & 2.7040 $\times 10^{-4}$ & 9.9305 $\times 10^{-4}$ & 0.031524 \tabularnewline
\hline
$g_1^u$ & -3.2055 & 0.92359 & -5.4828 & 9.4143 & -4.7444 & 2.8147 $\times 10^{-5}$ & 6.1713 $\times 10^{-3}$ & 0.064530 \tabularnewline
\hline
$g_1^d$ & 2.0946 & 1.0255 & -4.9812 & 7.5169 & -3.5011 & 9.6247 $\times 10^{-5}$ & 1.1900 $\times 10^{-3}$ & 0.072382 \tabularnewline
\hline
\hline
$\begin{aligned}
x_0 &= 0.2 \\
P^z &= 2.94 \; \mathrm{GeV}
\end{aligned}$
& $p_1$ & $p_2$ & $p_3$ & $p_4$ & $p_5$ & $\chi^{2}$ & $\overset{\circ}{r}$ & $\tilde{r}$ \tabularnewline
\hline
$f_1^u$ & -2.7310 & 1.1102 & -6.0771 & 12.308 & -6.6960 & 3.1037 $\times 10^{-5}$ & 9.1975 $\times 10^{-4}$ & 9.0825 $\times 10^{-3}$ \tabularnewline
\hline
$f_1^d$ & -2.8954 & 2.8391 & -6.0526 & 11.436 & -6.0637 & 8.605 $\times 10^{-4}$ & 5.3361 $\times 10^{-4}$ & 2.7550 $\times 10^{-3}$ \tabularnewline
\hline
$g_1^u$ & -2.7305 & 1.1882 & -5.5531 & 9.5354 & -4.5168 & 1.8665 $\times 10^{-8}$ & 6.7572 $\times 10^{-4}$ & 0.010776 \tabularnewline
\hline
$g_1^d$ & -1.7285 & 2.1573 & -4.1638 & 3.9141 & -0.055601 & 3.3523 $\times 10^{-6}$ & 4.7192 $\times 10^{-4}$ & 6.2346 $\times 10^{-3}$ \tabularnewline
\hline
\end{tabular}
%\par\end{centering}
\caption{\label{tab:rcnstrct_prms_x02} Numerical values of the reconstruction parameters in Eq.~\eqref{eq:param} and of the $\chi^2$ in Eq.~\eqref{eq:chi2} for all channels at the matching point $x_0 = 0.2$. Upper columns for $P^z = 1.47$ GeV, lower columns for $P^z = 2.94$ GeV. The $\overset{\circ}{r}$ and $\tilde{r}$ values represent the relative distance of the reconstructed PDFs $\overset{\circ}{q} (x, P^z)$ and quasi-PDFs $\tilde{q} (x, P^z)$ with respect to the standard PDFs $q(x)$, as defined in Eqs.~\eqref{eq:hatr} and \eqref{eq:tilder}, respectively.}
\end{table}

%%%%%%%%%%%%%%%%%%%%%%%%%%%%%%%%%%%%%%

\end{center}
\end{widetext}
%%%%%%%%

%%%%%% Fig. 4 - reconstruction at x0=0.2, Pz=2.94 GeV
%%%%%%%%%%%%%
%\vspace{-0.24cm}

The qualitative comments about the results of Figs.~\ref{fig:cmprsn_PDFs_x02Pz147},~\ref{fig:cmprsn_PDFs_x02Pz294} can be made more quantitative by looking at Tab.~\ref{tab:rcnstrct_prms_x02}. In this table, we list the values of the parameters of $\hat{q} (x, \left\{ p_i \right\})$ in Eq.~\eqref{eq:param} and of $\chi^2$ in Eq.~\eqref{eq:chi2} for all cases at $x_0 = 0.2$. In the last two columns, we show the numeric results for
\begin{align}
\overset{\circ}{r} [\overset{\circ}{q}] &= \frac{\int_0^1 dx \, [\overset{\circ}{q} (x, P^z) - q (x)]^2}{\int_0^1 dx \, q (x)^2} \, , \label{eq:hatr} \\
\tilde{r} [\tilde{q}] &= \frac{\int_0^1 dx \, [\tilde{q} (x, P^z) - q (x)]^2}{\int_0^1 dx \, q (x)^2} \, ,
\label{eq:tilder}
\end{align}
namely for the relative distances $\overset{\circ}{r}$ and $\tilde{r}$ of the reconstructed PDF $\overset{\circ}{q} (x, P^z)$ and quasi-PDF $\tilde{q} (x, P^z)$ with respect to the standard PDF $q (x)$, respectively. The values of $\tilde{r}$ quantify the level of agreement between the dashed lines (quasi-PDFs) and black solid lines (standard PDFs) shown in Figs.~\ref{fig:cmprsn_PDFs_x02Pz147}, \ref{fig:cmprsn_PDFs_x02Pz294}. When increasing $P^z$, the relative distance drops approximately by one order of magnitude except for the helicity $g_1^u$ of up quarks. The quality of our reconstruction procedure can be assessed through the definition of the relative distance $\overset{\circ}{r}$. We notice that the values of $\overset{\circ}{r}$ are systematically lower by one order of magnitude than the ones of $\tilde{r}$, sometimes by more as in the case $\tilde{g}_1^u (x, P^z = 2.94 \; \mathrm{GeV})$. Moreover, we can specify how much our procedure becomes more reliable when increasing $P^z$ by comparing the different values of $\overset{\circ}{r}$ for $P^z = 1.47$ GeV and $P^z = 2.94$ GeV: the reduction factor in the distance is larger than 2, and reaches one order of magnitude for the $\hat{g}_1^u$ channel.

In Fig.~\ref{fig:r_Pz_run}, we analyze in more detail the behaviour of the relative distance $\overset{\circ}{r}$ for different $P^z$ at the matching point $x_0 = 0.2$. In the left panel, filled diamonds connected by a dark (black) solid line represent $\overset{\circ}{r}$ for the unpolarized PDF of up quarks $f_1^u$, filled circles connected by a lighter (blue) solid line correspond to $f_1^d$, open diamonds connected by a long-dashed (black) line correspond to the helicity PDF of up quarks $g_1^u$, open circles connected by a short-dashed (blue) line correspond to $g_1^d$. In the right panel, the ratio between the distance $\overset{\circ}{r}$ of the reconstructed PDF $\overset{\circ}{q} (x, P^z)$ and the distance $\tilde{r}$ of the quasi-PDF $\tilde{q} (x, P^z)$ is shown as a function of $P^z$ with the same notation and in the same conditions as in the left panel. In terms of absolute values, the relative distance $\overset{\circ}{r}$ of the various reconstructed PDFs $\overset{\circ}{q} (x, P^z)$ is always very small, below 1\%, and for $P^z \gtrsim 2$ GeV it improves by almost one order of magnitude reaching the 0.1\% level for all channels. We note that for the down quark this very good level of accuracy is practically achieved for all the explored $P^z$ values. From the right panel, we deduce that for moderate $P^z$ the level of accuracy reached by our reconstruction procedure is more than ten times higher than for the quasi-PDFs. But when $P^z$ increases above 2 GeV, the quasi-PDFs become a good approximation to the standard PDFs (see also Fig.~\ref{fig:cmpr_LC_QS}): the relative distance $\tilde{r} [\tilde{q}]$ becomes smaller, and the ratio $\overset{\circ}{r} [\overset{\circ}{q}] / \tilde{r} [\tilde{q}]$ increases. This is particularly evident for $f_1^d$, described by the solid circles connected by the lighter (blue) solid line.

%%%%%%%%%
\begin{widetext}
\begin{center}

%%%%%% Fig. 5 - relative distances at x0=0.2
\vspace{-0.7cm}
\begin{figure}[!hbt]
\begin{center}
\includegraphics[width=0.42\textwidth]{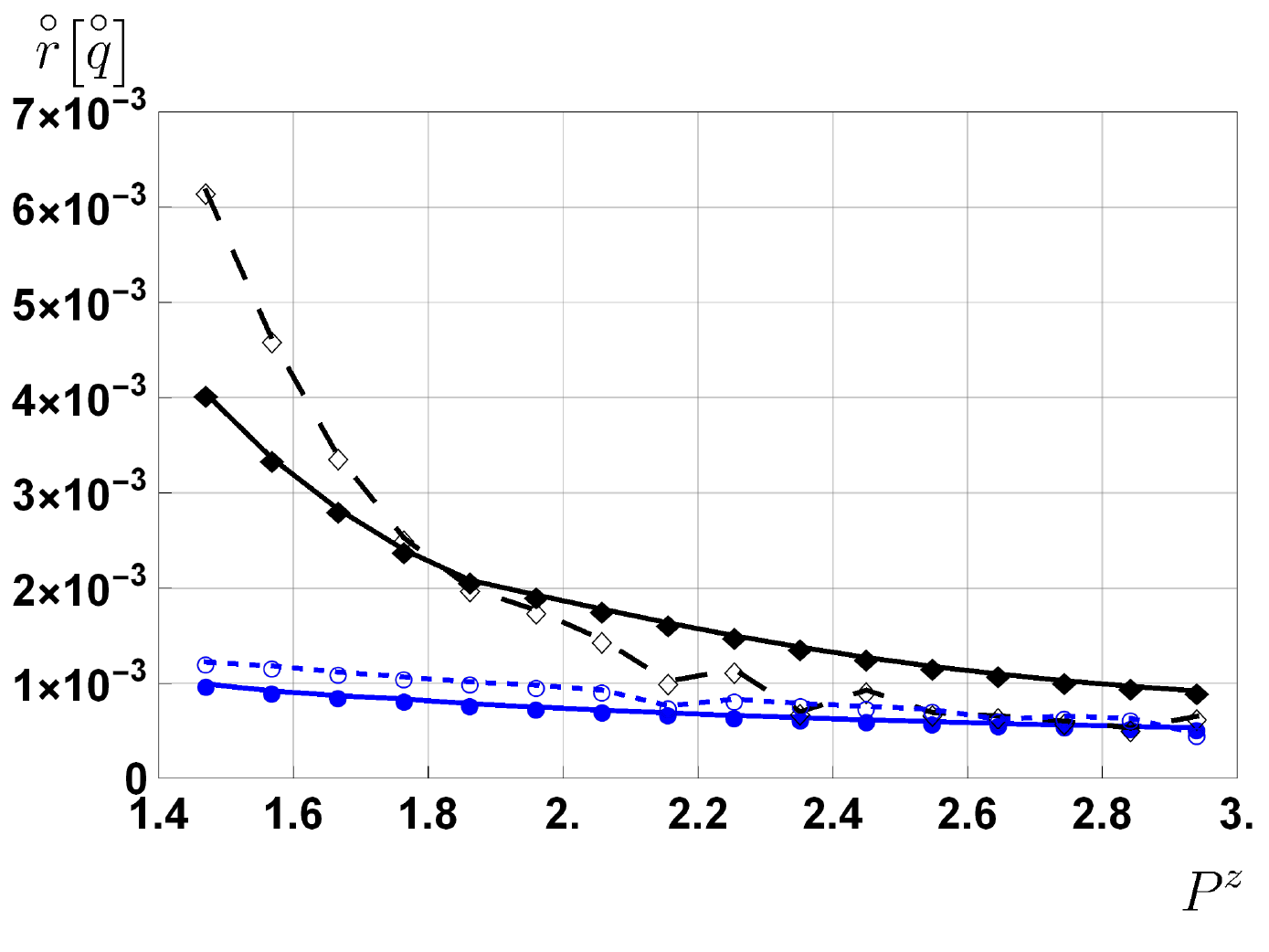}  \hspace{0.2cm}
\includegraphics[width=0.42\textwidth]{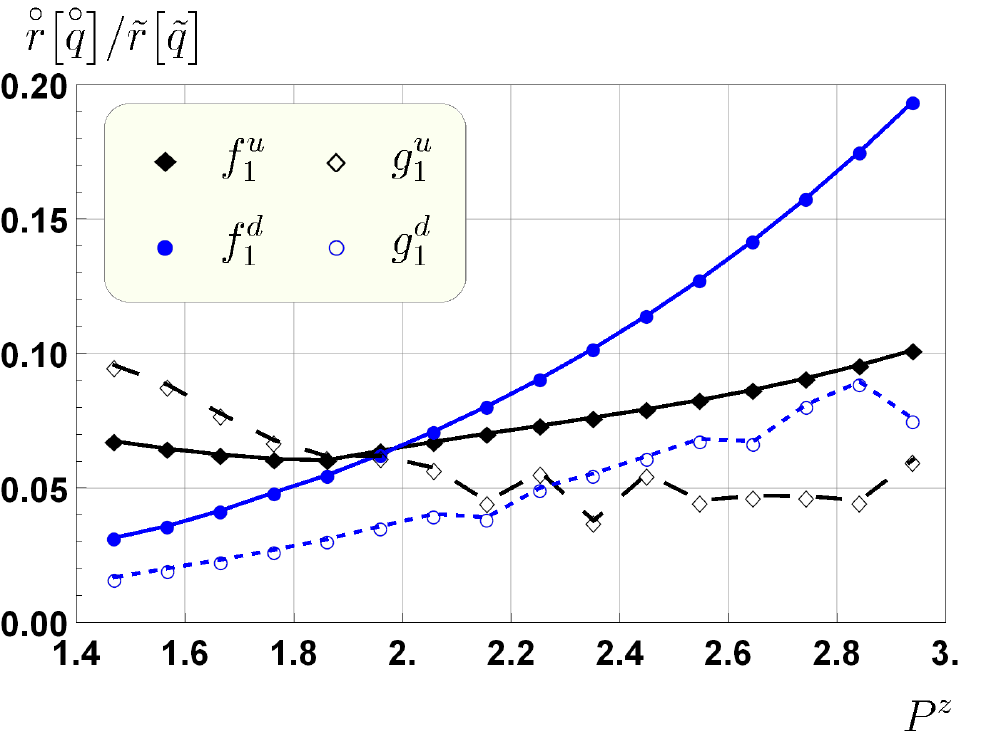}
\end{center}
\vspace{-0.5cm}
\caption{\label{fig:r_Pz_run} Left panel: relative distances $\overset{\circ}{r}$ for the various reconstructed PDFs $\overset{\circ}{q} (x, P^z)$ as functions of $P^z$ at $x_0 = 0.2$. Right panel: ratio of the $\overset{\circ}{r}$ distance with respect to the $\tilde{r}$ distance for the corresponding quasi-PDFs $\tilde{q} (x, P^z)$ as functions of $P^z$. Filled diamonds for the unpolarized PDF $f_1^u$, filled circles for $f_1^d$, open diamonds for the helicity PDF $g_1^u$, open circles for $g_1^d$.}
\end{figure}
%%%%%%%%%%%%%

\end{center}
\end{widetext}
%%%%%%%%

%%%%%% Fig. 6 -reconstruction at x0=0.3, Pz=1.47 GeV
\begin{figure}[!hbt]
\begin{center}
\includegraphics[width=0.48\textwidth]{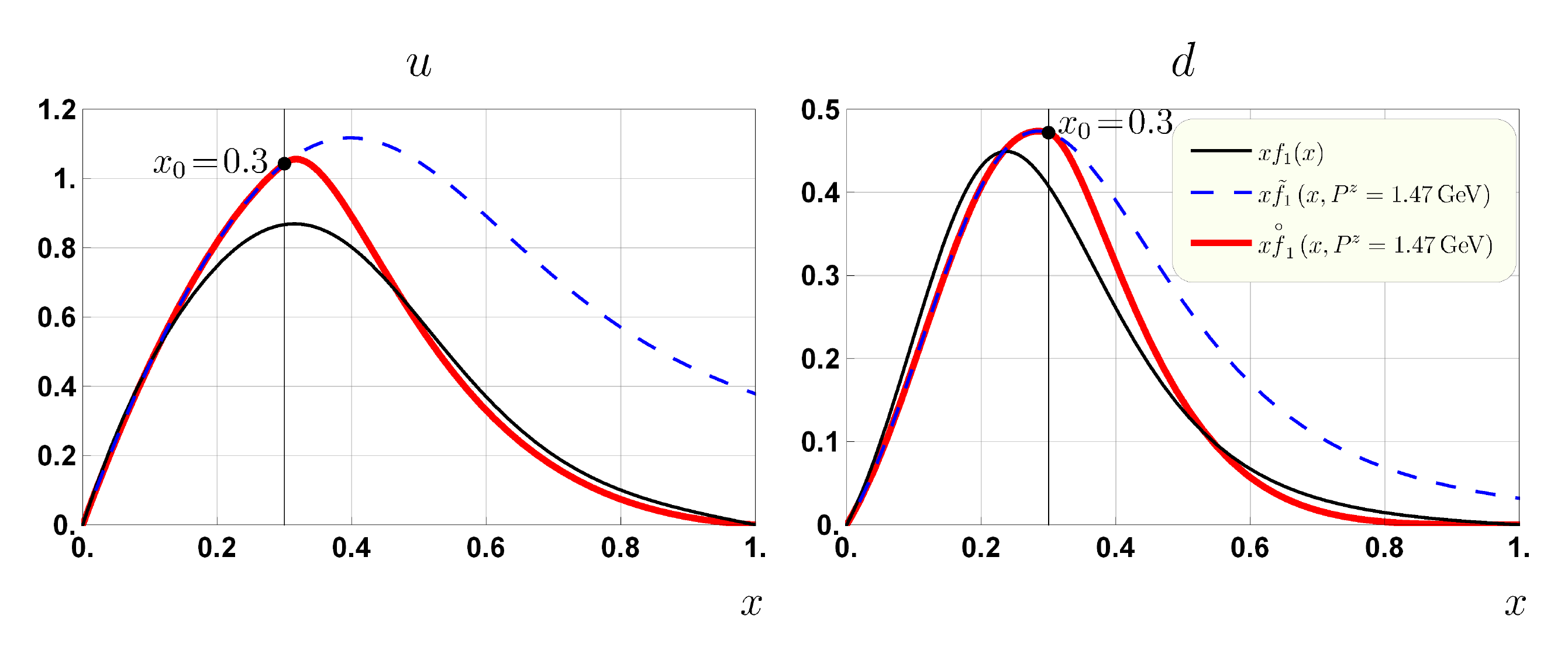} \\
\includegraphics[width=0.48\textwidth]{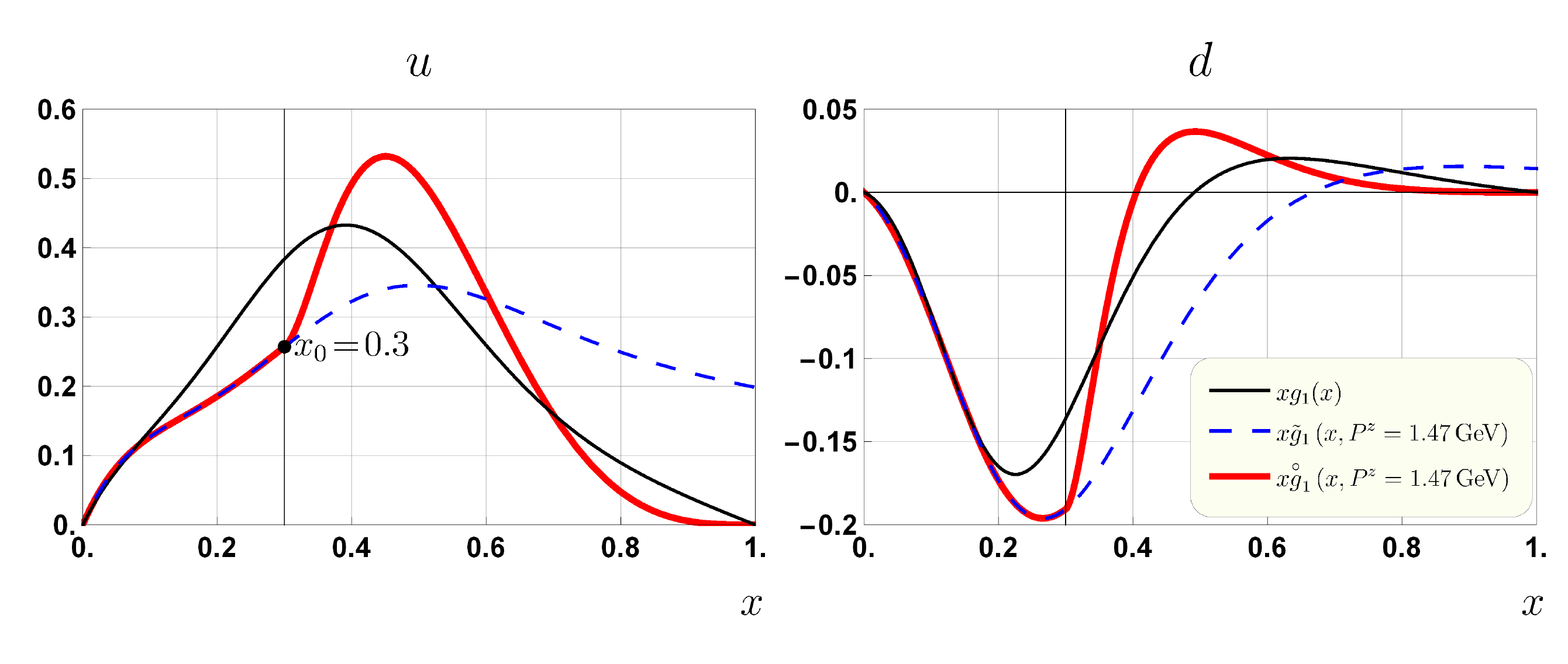}
\end{center}
\vspace{-0.5cm}
\caption{\label{fig:cmprsn_PDFs_x03Pz147} Same content, notation and conventions, as in Fig.~\ref{fig:cmprsn_PDFs_x02Pz147} but for $x_0 = 0.3$.}
\end{figure}
%%%%%%%%%%%%%

In Fig.~\ref{fig:cmprsn_PDFs_x03Pz147}, we display the comparison between standard PDFs, quasi-PDFs, and reconstructed PDFs, in the same conditions and notation as in Fig.~\ref{fig:cmprsn_PDFs_x02Pz147} but for the matching point $x_0 = 0.3$. It is evident that moving $x_0$ to higher values produces a worse situation: large oscillations in the reconstructed PDF $\overset{\circ}{q} (x, P^z)$ deteriorate the agreement with the standard PDF $q(x)$, particularly for the $q(x) \equiv g_1^u (x)$ case. This qualitative impression is confirmed by checking the numerical values at $P^z = 1.47$ GeV of the relative distances $\hat{r}$ against $\tilde{r}$ in Tab.~\ref{tab:rcnstrct_prms_x03}. While the $\tilde{r}$ are very similar to the corresponding numbers in Tab.~\ref{tab:rcnstrct_prms_x02}, the $\overset{\circ}{r}$ are almost one order of magnitude larger (except for the $f_1^u$ channel).

%%%%%% Fig. 7 - reconstruction at x0=0.3, Pz=2.94 GeV
\begin{figure}[!hbt]
\begin{center}
\includegraphics[width=0.48\textwidth]{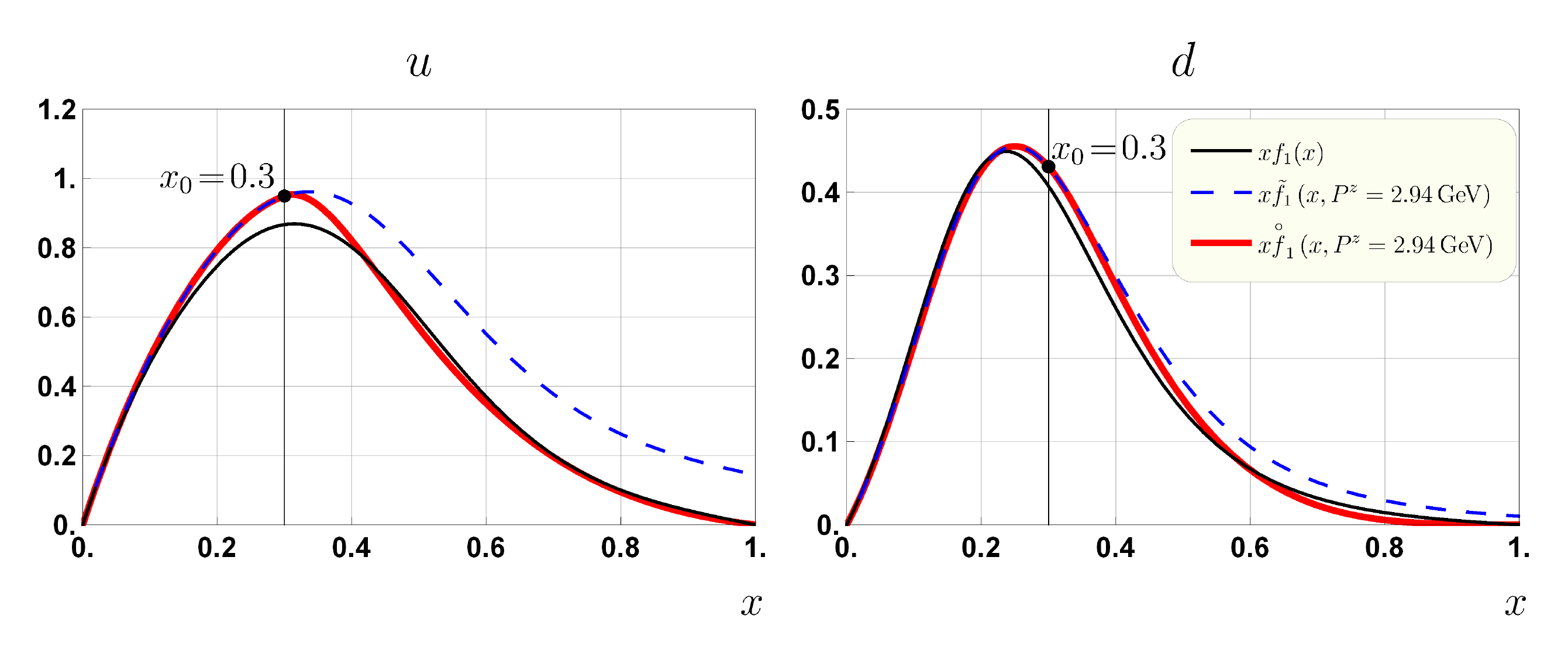}  \\
\includegraphics[width=0.48\textwidth]{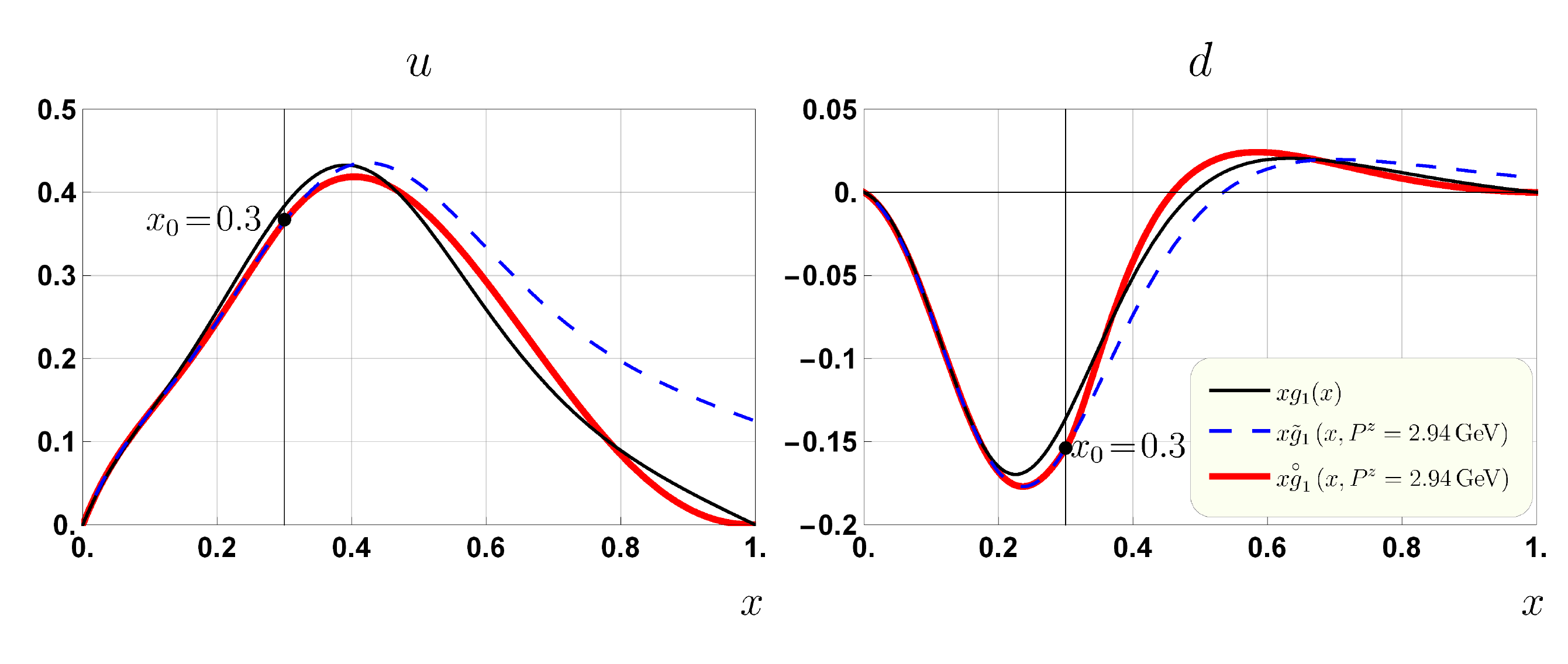}
\end{center}
\vspace{-0.5cm}
\caption{\label{fig:cmprsn_PDFs_x03Pz294} Same content, notation and conventions, as in Fig.~\ref{fig:cmprsn_PDFs_x02Pz294} but for $x_0 = 0.3$.}
\end{figure}
%%%%%%%%%%%%%

The overall accuracy of the reconstruction improves by moving to $P^z = 2.94$ GeV as displayed in Fig.~\ref{fig:cmprsn_PDFs_x03Pz294}, where the comparison is depicted again in the same conditions and notation as in Fig.~\ref{fig:cmprsn_PDFs_x02Pz294} but for the matching point $x_0 = 0.3$. This is confirmed by the values in Tab.~\ref{tab:rcnstrct_prms_x03} for $P^z = 2.94$ GeV: the $\hat{r}$ values are now similar or slightly larger than the ones in Tab.~\ref{tab:rcnstrct_prms_x02}, except for the $f_1^d$ channel.

We deduce that increasing $P^z$ is beneficial in various respects, but the best accuracy of our reconstruction procedure is reached for the lowest matching point $x_0 = 0.2$ because for $x \lesssim x_0$ the quasi-PDFs are a very good approximation to the standard PDFs. Although in this work we have computed the PDFs at the scale of the diquark spectator model, there is no restriction on applying the reconstruction procedure at higher scales provided that the weighted quasi-PDFs $x^{n-1} \tilde{q} (x, P^z)$ are a good approximation to the corresponding weighted standard PDFs $x^{n-1} q (x)$.

We conclude the section by testing how robust is our reconstruction procedure. To this aim, we perturb the various inputs to our procedure and we check how much the reconstructed PDF changes with respect to the unperturbed solution. More specifically, we shift by a certain amount $\delta$ the values of the quasi-PDF and of its first derivative at the matching point, 
\begin{align}
\tilde{q} (x_0, P^z) &\to (1+\delta) \tilde{q} (x_0, P^z) \nonumber \\
\frac{d}{dx} \tilde{q} (x, P^z) \Big\vert_{x = x_0} &\to (1+\delta) \frac{d}{dx} \tilde{q} (x, P^z) \Big\vert_{x = x_0} \, ,
\label{eq:shift}
\end{align}
as well as the difference between the (truncated) Mellin moments of the quasi-PDF and standard PDF, 
\begin{align}
\tilde{q}^n (P^z) - q^n &\to (1+\delta) \left( \tilde{q}^n (P^z) - q^n \right) \, ,
\label{eq:shiftMellin}
\end{align}
where $\tilde{q}^n (P^z), \  q^n$ are defined in Eq.~\eqref{eq:Mellin}. The minimization will produce a parametric expression $\hat{q}$ with new parameters $\{ p'_1, p'_2, p'_5 \}$. Namely, the weighted square distance of Eq.~\eqref{eq:chi2} becomes 
\begin{align}
\chi^{2} \left( \left\{ p'_1, p'_2, p'_5 \right\} \right) &= \nonumber \\
&\hspace{-1cm} \sum_{n=2}^4 \frac{\big[ \hat{q}^n \left( \left\{ p'_1, p'_2, p'_5 \right\} \right) + (1+\delta )\, ( \tilde{q}^{\, n} ( P^z ) - q^{n} ) \big]^2}{(1+\delta )^2 \, \big[ \tilde{q}^{\, n} \left( P^z \right) - q^n \big]^2} \, .
\label{eq:chi2prime}
\end{align}
The goal is to understand how much the parameters $\{ p'_1, p'_2, p'_5\}$ differ from the unperturbed ones $\{ p_1, p_2, p_5\}$. We define the perturbed parametric expression as 
\begin{align}
\hat{q} (x, \delta) &= \hat{q} (x, \{ p'_1, p'_2, p'_5 \} ) \, ,
\label{eq:hatqshift}
\end{align}
such that the unperturbed one is 
\begin{align}
\hat{q} (x, 0) &= \hat{q} (x, \{ p_1, p_2, p_5 \} ) \, .
\label{eq:hatqnoshift}
\end{align}
The robustness is measured by the relative distance between $\hat{q} (x, \delta)$ and $\hat{q} (x, 0)$ in the region $x_0< x \leq 1$, 
\begin{align}
\mathrm{r}(\delta) &= \frac{\int_{x_0}^1 dx \left[ \hat{q} (x, \delta) - \hat{q} (x, 0) \right]^2}
{\int_{x_0}^1 dx \left[ \hat{q} (x,0) \right]^2} \, .
\label{eq:robust}
\end{align}
The smaller $\mathrm{r} (\delta)$, the more stable the parametric expression $\hat{q} (x, \{ p_i \})$ in Eq.~\eqref{eq:QPDFdef}, the more robust the procedure leading to the reconstructed PDF $\overset{\circ}{q} (x, P^z)$. 

We perform the test with $P^z = 1.47$ GeV and $x_0 = 0.2$. In order to keep $\mathrm{r} (\delta) \lesssim 1$\%, for the unpolarized PDF of the up quark, $f_1^u$, we deduce $\left| \delta \right| \leq 0.09$, while for all the other channels we have $\left| \delta \right| \leq 0.1$. In other words, if we perturb the inputs by at most 10\%, the reconstructed PDF changes by no more than 1\%. This uncertainty is completely negligible with respect to the sensitivity of the reconstructed PDF when varying $P^z$ or $x_0$. Therefore, all the numerical results and related comments reported in Tabs.~\ref{tab:rcnstrct_prms_x02} and~\ref{tab:rcnstrct_prms_x03} are stable, solid and reliable. 

%%%%%%%%%%%
\begin{widetext}
\begin{center}

%%%%%%%%%% Tab. II - parameters and one-loop values at x0=0.3

\begin{table}[h]
\begin{tabular}{|c|c|c|c|c|c|c|c|c|}
\hline
$\begin{aligned}
x_0 &= 0.3 \\
P^z &= 1.47 \; \mathrm{GeV}
\end{aligned}$
& $p_1$ & $p_2$ & $p_3$ & $p_4$ & $p_5$ & $\chi^2$ & $\overset{\circ}{r}$ & $\tilde{r}$ \tabularnewline
\hline
\hline
$f_1^u$ & -4.0460 & 1.1645 & -5.5294 & 9.5218 & -4.7931 & 9.2887 $\times 10^{-3}$ & 3.5681$\times 10^{-3}$ & 0.059932 \tabularnewline
\hline
$f_1^d$ & -3.8033 & 3.1242 & -5.7196 & 10.069 & -5.1042 & 3.8004 $\times 10^{-3}$ & 0.016389 & 0.031524 \tabularnewline
\hline
$g_1^u$ & -3.0153 & 3.2889 & 9.2000 & -44.970 & 45.797 & 4.0819 $\times 10^{-4}$ & 0.018928 & 0.064530 \tabularnewline
\hline
$g_1^d$ & -2.0826 & 0.78932 & -5.0404 & 7.7482 & -3.7083 & 6.0952 $\times 10^{-2}$ & 0.032486 & 0.072382 \tabularnewline
\hline
\hline
$\begin{aligned}
x_0 &= 0.3 \\
P^z &= 2.94 \; \mathrm{GeV}
\end{aligned}$
& $p_1$ & $p_2$ & $p_3$ & $p_4$ & $p_5$ & $\chi^2$ & $\overset{\circ}{r}$ & $\tilde{r}$ \tabularnewline
\hline
$f_1^u$ & -3.8259 & 0.96646 & -5.4718 & 9.3989 & -4.7351 & 2.1711 $\times 10^{-3}$ & 6.3984 $\times 10^{-4}$ & 9.0825 $\times 10^{-3}$ \tabularnewline
\hline
$f_1^d$ & -3.3846 & 2.7794 & -5.6406 & 9.9664 & -5.0798 & 2.4672 $\times 10^{-3}$ & 2.8114 $\times 10^{-3}$ & 2.7550 $\times 10^{-3}$ \tabularnewline
\hline
$g_1^u$ & -1.4889 & 2.2966 & -5.5358 & 7.6106 & 1.2706 & 2.696 $\times 10^{-5}$ & 9.4131 $\times 10^{-4}$ & 0.010776 \tabularnewline
\hline
$g_1^d$ & -3.5531 & 1.5089 & -4.4239 & 6.1881 & -2.7115 & 1.0187 $\times 10^{-2}$ & 3.6958 $\times 10^{-4}$ & 6.2346 $\times 10^{-3}$ \tabularnewline
\hline
\end{tabular}
\caption{\label{tab:rcnstrct_prms_x03} Numerical values of the reconstruction parameters in Eq.~\eqref{eq:param} and of the $\chi^2$ in Eq.~\eqref{eq:chi2} for all channels at the matching point $x_0 = 0.3$. Upper columns for $P^z = 1.47$ GeV, lower columns for $P^z = 2.94$ GeV. The $\overset{\circ}{r}$ and $\tilde{r}$ values have the same meaning as in Tab.\ref{tab:rcnstrct_prms_x02}.}
\end{table}

%%%%%%%%%%%%%%%%%%%%%%%%%%%%%%%%%%%%%%%%%%%%%%%%

\end{center}
\end{widetext}
%%%%%%%%

%%%%%%%%%%%%%%%%%%%%%%%%%%%%%%%%%%%%%%%%%

\section{Conclusions}
\label{sec:end}

In this paper, we have presented a method to reconstruct a Parton Distribution Function (PDF) by combining information from its Mellin moments and from the corresponding quasi-PDF. Quasi-PDFs are obtained from hadronic matrix elements of equal-time spatial correlation operators; as such, they can be calculated on an Euclidean lattice. Quasi-PDFs can be  shown to reduce to the usual PDFs when the longitudinal momentum of the parent hadron becomes very large, in the limit $P^z \to \infty$. Lattice calculations of proton quasi-PDFs are already available but only for $P^z$ of the order of the proton mass, because for larger $P^z$ the computational effort is too demanding. Model calculations of quasi-PDFs are available in the framework of the diquark spectator approximation. They show that for moderate $P^z$ the quasi-PDFs are a good approximation to PDFs only for intermediate partonic momentum fractions $0.1 \lesssim x \lesssim 0.4$.

Our reconstruction procedure consists in choosing a matching point $x_0$,  and in merging the information delivered by the quasi-PDF at $0 \leq x \leq x_0$ with a parametric expression at $x_0 < x \leq 1$ that fits the $n = 2, 3, 4$ Mellin moments of the PDF itself. The minimization is constrained by requiring that at $x_0$ the quasi-PDF and the parametric expression match, including their first derivative. Since lattice calculations of quasi-PDFs at sufficiently large $P^z$ are missing, we have tested our method by using the diquark spectator approximation for up and down valence distributions of both unpolarized and helicity PDFs. We have also explored how the results change when varying $P^z$ at two different matching points $x_0 = 0.2$ and $0.3$.

In order to quantify the level of accuracy of our reconstruction, we have defined a normalized relative distance $\overset{\circ}{r}$ of the reconstructed PDF with respect to the standard PDF, and we have compared it with a similarly defined distance $\tilde{r}$ for the quasi-PDF. At $x_0 = 0.2$, the $\overset{\circ}{r}$ is always below 1\%, and for $P^z \gtrsim 2$ GeV it reaches the 0.1\% level for all channels. For the down quark, this very good level of accuracy is practically achieved for all the explored $P^z$ values. For $P^z \lesssim 2$ GeV, the $\overset{\circ}{r}$ is between ten and twenty times smaller than $\tilde{r}$: thus, at the $P^z$ currently reachable on lattice our reconstruction procedure reproduces the standard PDF in a much more reliable way than the quasi-PDF. When $P^z$ increases above 2 GeV, the quasi-PDFs also become a good approximation to the standard PDFs. But at $P^z = 3$ GeV the $\overset{\circ}{r}$ is still five times smaller than $\tilde{r}$ for the unpolarized up PDF $f_1^u$, and even smaller for the down quark and for the helicity PDF. At the matching point $x_0 = 0.3$, the general accuracy of the method deteriorates. In particular, the values of $\overset{\circ}{r}$ are very sensitive to the scale $P^z$. Only for $P^z \sim 3$ GeV they are similar to the ones reached at $x_0 = 0.2$. At the lower $P^z = 1.47$ GeV, we find that the $\overset{\circ}{r}$ at $x_0 = 0.3$ are approximately one order of magnitude larger than at $x_0 = 0.2$, with the only exception of the $f_1^u$ channel.

In summary, our method allows one to reconstruct a standard PDF by using the corresponding quasi-PDF up to $x_0 \simeq 0.2$ for values of $P^z$ as low as $1.5$ GeV, and then by fitting only few Mellin moments of the PDF itself. In this work, we have tested our approach using the results of a diquark spectator model at its natural low scale. However, our method can be used to obtain PDFs based on lattice QCD calculations at higher scales, where computations of Mellin moments of PDFs and their corresponding quasi-PDFs will be available.

%%%%%%%%%%%%%%%%%%%%%%%%%%%%%%%%%%%%%%%%%

\appendix

\begin{widetext}

\section{}
\label{sec:A}

Here below, we display the analytic expression of the integrated collinear quasi-PDFs in the spectator diquark model for both unpolarized and helicity distributions. In both cases, we have a scalar $(s)$ and an axial-vector $(a)$ components. They depend on the longitudinal parent nucleon momentum, $P^z$, on the fractional partonic momentum along the same direction, $x = k^z / P^z$, and on the diquark model parameters: the diquark mass $M_X$, the nucleon-quark-diquark coupling $g_X$, and the cutoff $\Lambda_X$ on the parton virtuality. The expressions are valid for $x$ in the range $[0,1[$, because the helicity quasi-PDFs are divergent in the $x=1$ point. 

%%%%

\subsection{Unpolarized PDF}
\label{sec:Af1}

The scalar-diquark component of the unpolarized collinear quasi-PDF is 
\begin{align}
\tilde{f}_1^s (x, P^z) &= 
\frac{g_s^2}{96 \pi ^2 P^0 \left( \Lambda_s ^2 - M^2 + 2 P^0 \sqrt{M_s^2 + (1-x)^2 (P^z)^2} - M_s^2 - 2 (P^z)^2 (1-x)\right)^3}
\nonumber \\
&\times \Bigg[ P^z \Big[ \Lambda_s ^2 + 2 m^2 + 4 m M x + x \left( 5 M^2 - \Lambda_s ^2 + M_s^2 + 6 (P^z)^2 (2-x) \right)   
                                         - 3 \left( M^2 + 2 (P^z)^2 + M_s^2 \right) \Big]
\nonumber \\
&\qquad            + 6 (1-x) P^0 P^z \sqrt{M_s^2 + (1-x)^2 (P^z)^2} \Bigg] \, ,
\label{eq:qf1sPDF}
\end{align}
where $P^0=\sqrt{(P^z)^2+M^2}$. 

The axial-vector-diquark component of the unpolarized collinear quasi-PDF is 
\begin{align}
\tilde{f}_1^a (x, P^z)  &= 
\frac{g_a^2}{384 \pi ^2 (P^0)^3 (P^z)^3 \left( M_a^2 + (P^z)^2 (1-x)^2 \right) \left( \Lambda_a^2 - M^2 + 2 P^0 \sqrt{M_a^2 + (P^z)^2 (1-x)^2} - M_a^2 - 2 (P^z)^2 (1-x) \right)^3} 
\nonumber \\
&\times \Bigg[ \Big[ 4 x (P^0)^2 (P^z)^4 \left( M^2 + M_a^2 + 2 (1-x) (P^z)^2 - \Lambda_a^2 \right) - 4 (P^0)^2 (P^z)^4 (1-x) \left( M_a^2 + (P^z)^2 (1-x)^2 \right) \Big] 
\nonumber \\
&\qquad \times \left( M^2 - 6 P^0 \sqrt{M_a^2 + (P^z)^2 (1-x)^2} + M_a^2 + 2 (1-x) (P^z)^2 - \Lambda_a^2 \right) 
\nonumber \\
&\quad - 8 (P^0)^2 (P^z)^4 \left( M_a^2 + (P^z)^2 (1-x)^2 \right) \Big[ M^2 + M_a^2 - m^2 - 2 m M x + 2 (P^z)^2 (1- (1-x) x)  - 6 x (P^0)^2 \Big]  \Bigg] \, .
\label{eq:qf1aPDF}
\end{align}

%%%%

\subsection{Helicity PDF}
\label{sec:Ag1}

The scalar-diquark component of the helicity collinear quasi-PDF is 
\begin{align}
\tilde{g}_1^s (x, P^z) &= 
- \frac{g_s^2}{96 \pi^2 (P^0)^2 \left( M^2 - 2 P^0 \sqrt{M_s^2 + (1-x)^2 (P^z)^2} + M_s^2 + 2 (1-x) (P^z)^2 - \Lambda_s^2 \right)^3}
\nonumber \\
&\times \Bigg[ \Big( M (m + M) + (P^z)^2 (1-x) \Big) \left( M^2 - 6 P^0 \sqrt{M_s^2 + (1-x)^2 (P^z)^2} + M_s^2 + 2 (1-x) (P^z)^2 - \Lambda_s^2 \right) 
\nonumber \\
&\qquad + 2 (P^0)^2 \Big( M_s^2 + (m + M)^2 + 2 (P^z)^2 (1-x)^2 \Big) \Bigg] \, .
\label{eq:qg1sPDF}
\end{align}

The axial-vector-diquark component of the helicity collinear quasi-PDF is 
\begin{align}
\tilde{g}_1^a (x, P^z) &= \nonumber \\
&\hspace{-1cm} - \frac{g_a^2}{384 \pi^2 (P^0)^3 \Big( M_a^2 + (P^z)^2 (1-x)^2 \Big) \left( M^2 - 2 P^0 \sqrt{M_a^2 + (1-x)^2 (P^z)^2} + M_a^2 + 2 (1-x) (P^z)^2 - \Lambda_a^2 \right)^3}
\nonumber \\
&\hspace{-1cm} \times \Bigg[ \  8 (P^0)^3 \Big( M_a^2 + (P^z)^2 (1-x)^2 \Big) \Big( 2 (P^z)^2 (1+x)^2 - M_a^2 - (m+M)^2 - 4 m M \Big) 
\nonumber \\
&\hspace{-0.5cm} + 4 P^0 \Big( M^2 - 6 P^0 \sqrt{M_a^2 + (1-x)^2 (P^z)^2} + M_a^2 + 2 (1-x) (P^z)^2 - \Lambda_a^2 \Big) \nonumber \\
&\hspace{-0.2cm} \times \bigg( \Big( M^2 + M_a^2 + 2 (1-x) (P^z)^2 - \Lambda_a^2 \Big) \Big( x (P^z)^2 - m M \Big) 
\nonumber \\
&\quad \  - \Big( M (m + M) + (P^z)^2 (1-x) \Big) \Big( M_a^2 + (P^z)^2 (1-x)^2 \Big) \bigg) \Bigg] \, .
\label{eq:qg1aPDF}
\end{align}

As explained in Eqs.~\eqref{eq:uX} and~\eqref{eq:dX}, the axial-vector quasi-PDFs $\tilde{f}_1^a, \, \tilde{g}_1^a$ can be further distinguished to depend on model parameters for an isoscalar-axial-vector diquark $(a)$ or for an isovector-axial-vector diquark $(a')$. The former is combined with the isoscalar-scalar component $s$ to give the quasi-PDF for the up quark, the latter directly contributes to the down quark. Each diquark component of quasi-PDFs entering the linear combination needs to be properly normalized. As explained in Sec.~\ref{sec:qPDFplot}, this amounts to replace $g_X^2$ in the above formulae with $c_X^2 \, N^2_X$, with $X = s, a, a'$, where $c_X$ are fitting parameters of the model and 
\begin{align}
N^2_X &= \Bigg[ \int_0^1 dx \lim_{P^z \to \infty} \frac{1}{g_X^2} \tilde{f}_1^X (x, P^z; M_X, \Lambda_X, g_X) \Bigg]^{-1}
\label{eq:norm}
\end{align}
is the normalization of the quasi-PDF for the diquark component $X$ (with the dependence of the quasi-PDF on the diquark model parameters made explicit).  

\end{widetext}

%%%%%%%%%%%%%%%%%%%%%%%%%%%%%%%%%%%%%%%%%%%%%%%%%

\bibliographystyle{apsrevM}
\bibliography{mybiblio}

%%%%%%%%%%%%%%%%%%%%%%%%%%%%%%%%%%%%%%%%%%%%%%%%%%

\end{document}